\documentclass[useAMS,usedcolumn,usenatbib,usegraphicx]{mn2e}
\usepackage{times}

\usepackage{aas_macros}
\usepackage{textcmds}
\usepackage{multirow}
\usepackage{xcolor}
\usepackage{hyperref}
\usepackage{amsmath,amssymb}
\usepackage{diagbox, lscape, chemfig, subcaption}
\usepackage{xtab,afterpage}
\usepackage{changepage}
\newcommand\T{\rule{0pt}{2.6ex}}       
\newcommand\B{\rule[-1.2ex]{0pt}{0pt}} 

\DeclareSymbolFont{matha}{OML}{txmi}{m}{it}
\DeclareMathSymbol{\varv}{\mathord}{matha}{118}

\title[Prestellar grain-surf. origins: D-methanol 67P/C-G]{Prestellar grain-surface origins of deuterated methanol in comet 67P/Churyumov--Gerasimenko}

\author[Maria N. Drozdovskaya et al.]{Maria~N.~Drozdovskaya$^{1}$\thanks{E-mail: maria.drozdovskaya@csh.unibe.ch}, Isaac R.~H.~G. Schroeder, I$^{2}$, Martin Rubin$^{2}$,\newauthor
Kathrin Altwegg$^{2}$, Ewine F. van Dishoeck$^{3, 4}$, Beatrice M. Kulterer$^{1}$,\newauthor
Johan De Keyser$^{5}$, Stephen A. Fuselier$^{6, 7}$, Michael Combi$^{8}$\\
$^{1}$~Center for Space and Habitability, Universit\"{a}t Bern, Gesellschaftsstrasse 6, CH-3012 Bern, Switzerland\\
$^{2}$~Physikalisches Institut, Universit{\"a}t Bern, Sidlerstrasse 5, CH-3012 Bern, Switzerland\\
$^{3}$~Leiden Observatory, Leiden University, PO Box 9513, NL-2300 RA Leiden, The Netherlands\\
$^{4}$~Max-Planck-Institut f\"{u}r Extraterrestrische Physik, Giessenbachstrasse 1, D-85748 Garching, Germany\\
$^{5}$~Royal Belgian Institute for Space Aeronomy, Ringlaan 3, B-1180 Brussels, Belgium\\
$^{6}$~Southwest Research Institute, 6220 Culebra Road, San Antonio, TX 78228, USA\\
$^{7}$~University of Texas at San Antonio, 1 UTSA Circle, San Antonio, TX 78249, USA\\
$^{8}$~Department of Climate and Space Sciences and Engineering, University of Michigan, 2455 Hayward Street, Ann Arbor, MI 48109, USA
}

\begin{document}

\date{Accepted xxx.  Received xxx; in original form xxx}

\pagerange{\pageref{firstpage}--\pageref{lastpage}} \pubyear{2020}

\maketitle
\label{firstpage}

\begin{abstract}
Deuterated methanol is one of the most robust windows astrochemists have on the individual chemical reactions forming deuterium-bearing molecules and the physicochemical history of the regions where they reside. The first-time detection of mono- and di-deuterated methanol in a cometary coma is presented for comet 67P/Churyumov--Gerasimenko using \textit{Rosetta}--ROSINA data. D-methanol (CH$_{3}$OD and CH$_{2}$DOH combined) and D$_{2}$-methanol (CH$_{2}$DOD and CHD$_{2}$OH combined) have an abundance of $5.5\pm0.46$ and $0.00069\pm0.00014$ per cent relative to normal methanol. The data span a methanol deuteration fraction (D/H ratio) in the $0.71-6.6$ per cent range, accounting for statistical corrections for the location of D in the molecule and including statistical error propagation in the ROSINA measurements. It is argued that cometary CH$_{2}$DOH forms from CO hydrogenation to CH$_{3}$OH and subsequent H-D substitution reactions in CH$_{3}$-R. CHD$_{2}$OH is likely produced from deuterated formaldehyde. Meanwhile, CH$_{3}$OD and CH$_{2}$DOD, could form via H-D exchange reactions in OH-R in the presence of deuterated water ice. Methanol formation and deuteration is argued to occur at the same epoch as D$_{2}$O formation from HDO, with formation of mono-deuterated water, hydrogen sulfide, and ammonia occurring prior to that. The cometary D-methanol/methanol ratio is demonstrated to agree most closely with that in prestellar cores and low-mass protostellar regions. The results suggest that cometary methanol stems from the innate cold ($10-20$~K) prestellar core that birthed our Solar System. Cometary volatiles individually reflect the evolutionary phases of star formation from cloud to core to protostar.
\end{abstract}


\clearpage
\newpage
\section{Introduction}
\label{intro}

Isotopologues are a unique window on the assembly of a molecule. If a molecule undergoes neither destruction and reformation, nor internal restructuring, then the ratio of the isotopes and their locations in that molecule can be used to directly pinpoint its chemical formation routes and physical conditions at the time of its creation. This method requires firm constraints on the isotopic ratios and the formation mechanisms that prescribe the location of isotopes in a specific species. A popular choice in astrochemistry are deuterated molecules, which have been suggested to stem from the earliest, coldest prestellar stages of star formation \citep{vanDishoeck1995, CaselliCeccarelli2012, CeccarelliPPVI}. Formation of methanol is one of the most thoroughly studied and well-constrained chemical reaction networks among interstellar molecules \citep{WatanabeKouchi2002b, Osamura2004, Nagaoka2005, Fuchs2009, Hidaka2009}. Consequently, deuterated methanol is one of the most robust windows astrochemists have on the individual chemical reactions and the physical conditions that lead to the formation of volatiles that are found in star-forming regions and our Solar System's comets.

The elemental abundance of deuterium relative to hydrogen ($A_{\text{D}}$) in the local (within $\sim 1 - 2$~kpc of the Sun) interstellar medium (ISM) has been derived to be at least $\left( 2.0 \pm 0.1 \right) \times 10^{-5}$, while accounting for observed variations stemming from depletion of gas-phase deuterium on to dust grains \citep{Vidal-Madjar2002, Burles2002, Linsky2006, Prodanovic2010}. At the cold ($10-20$~K) temperatures of prestellar cores, the molecular deuteration fraction (called the D/H ratio of a molecule) of some species can be several orders of magnitude larger. Deuterium readily enters gas-phase and grain-surface chemical reactions at the temperatures, densities, and ionization fractions of cores by being liberated from HD by H$_{3}^{+}$ into H$_{2}$D$^{+}$ \citep{Watson1974, DalgarnoLepp1984, Caselli2019b}. At slightly higher densities ($\sim10^{4}$~cm$^{-3}$), gas-phase abundance of atomic D is thought to be higher than that of HD, allowing it to be directly incorporated into other gaseous and solid molecules \citep{Tielens1983}. Mono-deuterated methanol is now routinely observed in star-forming regions in the low- and high-mass regimes (e.~g., \citealt{Bogelund2018, Agundez2019, Manigand2020a}). Di- (in low-mass protostars IRAS~16293-2422, \citealt{Parise2002}, and SVS13-A, \citealt{Bianchi2017a}) and tri-deuterated (in IRAS~16293-2422, \citealt{Parise2004}) methanol have been also detected, indicating that efficient incorporation of deuterium into methanol occurs in star-forming regions.

In cometary comae, methanol is one of the major volatiles detected at a level of $0.2-7$ per cent relative to water \citep{MummaCharnley2011, LeRoy2015}. Deuterated molecules, such as HDO and DCN, have also been measured in a number of comets; however, deuterated methanol has not been identified thus far \citep{Bockelee-Morvan2015}. Recently, the ESA \textit{Rosetta} mission provided a unique opportunity to characterize the Jupiter Family Comet (JFC) 67P/Churyumov–Gerasimenko, hereafter 67P/C–G, at an unprecedented level of detail through in situ measurements during a two-year monitoring phase. It has been revealed that the volatiles of 67P/C--G are chemically diverse and complex; and that most isotopic ratios measured in the volatile species are non-Solar \citep{Hoppe2018}. Measurements of the D/H ratio of 67P/C--G in water, hydrogen sulfide, and ammonia show that the comet is enriched in the minor D isotope relative to the ISM isotopic ratio \citep{Altwegg2019}. These results have already been interpreted as indicators of a low-temperature formative scenario for JFCs \citep{Altwegg2015, Altwegg2017a}.

In this paper, the first-time detection of mono- and di-deuterated methanol (henceforth, also denoted by D-methanol and D$_{2}$-methanol, respectively) is presented in comet 67P/C–G, as obtained with the Rosetta Orbiter Spectrometer for Ion and Neutral Analysis (ROSINA; \citealt{Balsiger2007}) instrument aboard the ESA \textit{Rosetta} spacecraft. Section~\ref{meth} describes the methods of data analysis. Section~\ref{results} presents the mass spectra, the derived range of D/H ratios in methanol, and compares the abundance ratios to the full set of currently available observations of deuterated methanol in star-forming regions. The chemistry of deuterated methanol is discussed in Section~\ref{discussion} alongside the implications these findings have in the astrochemical context for the evolutionary sequence of star- and comet-forming regions. Section~\ref{discussion} also addresses the constraints placed by the presented measurements on the physical conditions of our Solar System's formation. The conclusions are summarized in Section~\ref{conclusions}.

\section{Methods}
\label{meth}


The ROSINA Double Focusing Mass Spectrometer (DFMS) has a high-mass resolution of $m / \Delta m = 3000$ for a $m/z$ (mass-to-charge ratio) of $28$ at $1$ per cent of the peak height \citep{Balsiger2007}. Its main detector, the MCP/LEDA, is a position-sensitive imaging detector that is a stack of two micro-channel plates (MCPs) arranged in Chevron configuration, followed by two independent rows (Rows A and B) of $512$ anodes on a linear electron detector array (LEDA). For the measurement mode discussed in this paper, spectra around each integer mass are obtained consecutively every $30$~s ($10$~s for adjusting voltages and $20$~s of integration time). The DFMS electron impact ionization source produces $45$~eV electrons that bombard and ionize the parent species, which ionize and/or fragment in a species-dependent characteristic fragmentation pattern \citep{DeKeyser2019b}. For a more complete description of the DFMS data analysis the reader is referred to \citet{LeRoy2015}, \citet{Calmonte2016}, and the references therein.

The overall gain (degree of amplification) produced by the MCP depends on which of the $16$ predefined voltage settings (gain steps) is applied. However, the gain corresponding to each gain step changed over time as the detector aged. This change had to be corrected for in all DFMS measurements \citep{Schroeder2019a}. An additional flat-field correction known as the ``pixel gain'' was also necessary, due to the non-uniform degradation of the $512$ LEDA anodes (pixels) caused by the uneven usage of the MCP. The appendix of \citet{Schroeder2019b} contains a full description of these corrections and how they were applied.

Each peak in a DFMS mass spectrum is best described by a double-Gaussian, where the second Gaussian has a peak height $a_{2}$ of approximately $10$ per cent that of the first one and a width $c_{2}$ that is roughly three times broader than the first \citep{DeKeyser2019a}:
\begin{equation}
f \left( x \right) = a_{1} e^{-\left( \frac{x - b}{c_{1}} \right)^{2}} + a_{2} e^{-\left( \frac{x - b}{c_{2}} \right)^{2}},
\end{equation}
where the parameters $a_{1}$ and $a_{2}$ are the amplitudes of the first and second Gaussians, respectively. Expressed in pixels are the peak centre, $b$, the corresponding widths, $c_{1}$ and $c_{2}$, and the variable $x$. The conversion of pixels to mass (as in, e.~g., Fig.~\ref{fig_ROSINA_32}) is described in \citet{Calmonte2016}. The peak widths, $c_{1}$ and $c_{2}$, and the amplitude ratio $a_{1}/a_{2}$ are kept constant for each peak in a single spectrum. The area encompassed by any given peak is the integral of its fitted double-Gaussian function:
\begin{equation}
\int^{\infty}_{-\infty} f \left( x \right) dx = \sqrt{\pi} \left( a_{1} c_{1} + a_{2} c_{2} \right).
\end{equation}

Minor isotopologues are best investigated at times of high outgassing rates of the main variant of that molecule and/or at close cometocentric distances of the orbiter. Several instances of high local methanol abundances during the monitoring phase of \textit{Rosetta} occurred early in the mission in October and December 2014. At those times, the orbiter was in close orbits at distances of $\sim10-30$~km above the comet. Due to aging of the detector over the course of the mission, this is the best period for fitting mass peaks associated with methanol and its isotopologues. Comet 67P/C--G was at heliocentric distances in the $3.2$ to $2.7$~au range during these months with the Northern hemisphere experiencing the summer season (inbound equinox was in May 2015; and outbound equinox was in March 2016) pre-perihelion (perihelion was on August 13th, 2015). 



\section{Results}
\label{results}

\subsection{67P/C--G: ROSINA mass spectra}
\label{mass_spec}

\begin{figure}
    \centering
    \begin{subfigure}[b]{0.45\textwidth}
        \includegraphics[width=\textwidth]{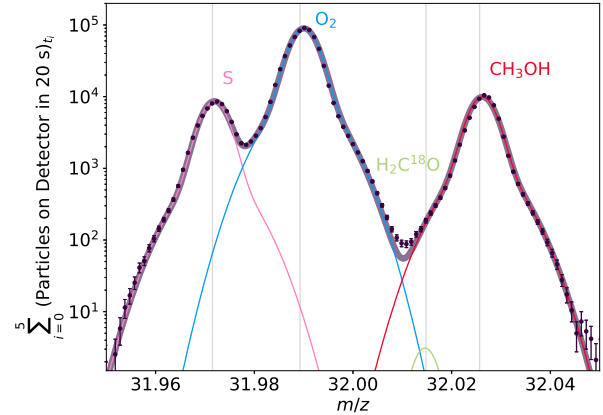}
        \caption{Normal methanol}
				\label{fig_ROSINA_32}
    \end{subfigure}
    ~ 
    \begin{subfigure}[b]{0.45\textwidth}
        \includegraphics[width=\textwidth]{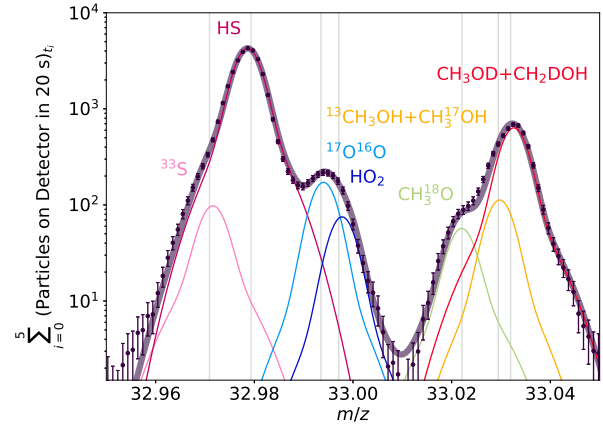}
        \caption{Mono-deuterated methanol}
				\label{fig_ROSINA_33}
    \end{subfigure}
		\begin{subfigure}[b]{0.45\textwidth}
        \includegraphics[width=\textwidth]{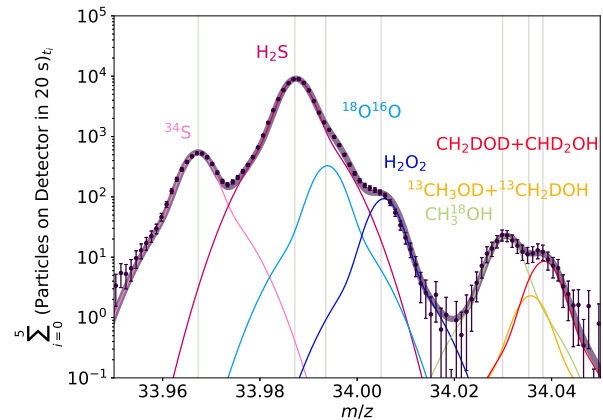}
        \caption{Di-deuterated methanol}
				\label{fig_ROSINA_34}
    \end{subfigure}
		\caption{ROSINA mass spectra for normal, mono-deuterated, and di-deuterated methanol on $m/z = 32, 33,$ and $34$, respectively, as measured on Row A and summed for the late October--December data set ($7$ in total: one packet on October 27th, three on December 9th, one on the 10th, and two on the 18th, 2014) are shown as the dark data points. The depicted statistical error bars are smaller than the data points themselves when they are not visible. The individual contributors to the mass peaks are indicated with vertical lines at their exact masses, and associated double Gaussian fits are shown as thin colored curves. The thick dark purple curve is the sum of the individual double Gaussians and represents the overall fit to the measured ROSINA mass spectra.}
\end{figure}

The mass spectra for normal (i.~e., non-deuterated) methanol, mono-deuterated methanol, and di-deuterated methanol at $m/z = 32$, $33$, and $34$, respectively, are investigated manually based on two data sets. One set matching the spectra presented by \citet{Altwegg2020b} for the study of oxygen isotopologues, which is a sum of three packets on October 9th and three on the 19th, 2014 ($6$ in total; i.~e., the early October data set, presented in Appendix~\ref{mass_spec_Oct14}). The second set is a sum of one packet on October 27th, three on December 9th, one on the 10th, and two on the 18th ($7$ in total; i.~e., late October--December data set Figs.~\ref{fig_ROSINA_32}--\ref{fig_ROSINA_34}). Only measurements on Row A are considered in both data sets (Section~\ref{meth}). The methanol peak on $m/z = 32$ is well-separated from the neighboring peaks associated with molecular oxygen and atomic sulphur, and consequently, can be easily fitted. The peak associated with H$_{2}$C$^{18}$O is weak and does not have a significant contribution to the methanol ion count. The mono-deuterated methanol peak on $m/z = 33$, on the other hand, has a strong overlap with peaks associated with the $^{13}$C- and $^{17}$O-bearing methanol isotopologues, and a fragment of the $^{18}$O-bearing methanol isotopologue. This overlap requires manual analysis of the mass spectrum. The peak centres are set by the precise masses of their corresponding species. The widths of the Gaussians are expected to be nearly identical on a single $m/z$ value. This width is deduced from the adjacent strong HS peak that is well-separated from its neighboring weaker $^{33}$S, $^{17}$O$^{16}$O, and HO$_{2}$ peaks. More specifically, it is imposed that every $c_{1}$ equals $c_{1}(\text{HS})$, every $c_{2}$ equals $c_{2}(\text{HS})$, and every $a_{1} / a_{2}$ equals $a_{1}(\text{HS}) / a_{2}(\text{HS})$ in this mass spectrum. This leaves the amplitude, $a_{1}$ or $a_{2}$, as the only free parameter, which can now be fitted separately for CH$_{3}^{18}$O, $^{13}$CH$_{3}$OH$ + $CH$_{3}^{17}$OH, and CH$_{3}$OD$ + $CH$_{2}$DOH peaks. For normal methanol, the width of the Gaussian is deduced from the strongest peak in that spectrum, which is that of O$_{2}$.

The mass spectra from early October and late October--December (Appendix~\ref{mass_spec_Oct14}; Fig.~\ref{fig_ROSINA_32} and~\ref{fig_ROSINA_33}) correspond to an average ratio of CH$_{3}$OD$ + $CH$_{2}$DOH to CH$_{3}$OH of $0.055\pm0.0046$ (henceforth, also denoted as D-methanol/methanol), where the error on the average of the two measurements from the two considered time intervals is calculated according to statistical error propagation. The abundance ratios relative to normal methanol for the early October and the late October--December data sets individually are given in Table~\ref{tabl_ROSINA_OC}. It is not possible to distinguish CH$_{3}$OD from CH$_{2}$DOH in the mass spectra from ROSINA at $m/z = 33$. The mass fragment CH$_{3}$O of CH$_{3}$OD cannot be distinguished from the CH$_{3}$O fragment stemming from CH$_{3}$OH (or the many other heavier O-bearing hydrocarbons). The mass fragment CH$_{2}$DO could stem from CH$_{2}$DOH upon loss of H from the hydroxyl (OH-R) functional group or from CH$_{3}$OD upon loss of H from the methyl (CH$_{3}$-R) functional group. Consequently, fragments also cannot aid in distinguishing the two mono-deuterated variants of methanol in the ROSINA data.


Fig.~\ref{fig_ROSINA_34} shows the mass spectrum for di-deuterated methanol for the late October--December data set (the early October data set is in Appendix~\ref{mass_spec_Oct14}). The di-deuterated methanol peak on $m/z = 34$ has a strong overlap with peaks associated with the $^{18}$O-bearing methanol isotopologue, and the double isotopologue bearing $^{13}$C and D. This overlap again requires manual analysis of the mass spectrum analogously to the analysis carried out at $m/z = 33$, but with the widths of the Gaussians and their amplitude ratio now deduced from the highest H$_{2}$S peak. The amplitude is then fitted separately for CH$_{3}^{18}$OH, $^{13}$CH$_{3}$OD + $^{13}$CH$_{2}$DOH, and CH$_{2}$DOD + CHD$_{2}$OH peaks. The double methanol isotopologue bearing $^{13}$C and D is thought to be rare, leaving the peak at $m/z = 34$ dominated by a separable overlap between CH$_{3}^{18}$OH and CH$_{2}$DOD + CHD$_{2}$OH \citep{Altwegg2020b}.

The mass spectra from early October and late October--December (Appendix~\ref{mass_spec_Oct14}; Fig.~\ref{fig_ROSINA_32} and~\ref{fig_ROSINA_34}) correspond to an average ratio of CH$_{2}$DOD$ + $CHD$_{2}$OH to CH$_{3}$OH of $0.00069\pm0.00014$ (henceforth, also denoted as D$_{2}$-methanol/methanol), where the error on the average of the two measurements from the two considered time intervals is calculated according to statistical error propagation. The abundance ratios relative to normal methanol for the early October and the late October--December data sets individually are given in Table~\ref{tabl_ROSINA_OC}. It is not possible to distinguish CH$_{2}$DOD from CHD$_{2}$OH in the mass spectra from ROSINA at $m/z = 34$. ROSINA data do not show a signal of CH$_{3}$D$^{18}$O, CHD$_{3}$O, nor CD$_{4}$O at $m/z = 35$, $36$, $37$, respectively.

\citet{Altwegg2020b} derived a $^{16}$O/$^{18}$O ratio of $495\pm40$ and a $^{12}$C/$^{13}$C ratio of $91\pm10$ for methanol based on the early October data set investigated in this work, upon assuming that $^{13}$CH$_{3}$OH$ \gg $CH$_{3}^{17}$OH. The mass spectra presented in this work are consistent within errors with these methanol oxygen and carbon isotopic ratios (Appendix~\ref{mass_spec_Oct14}). The measured methanol abundances relative to water and total production rates from ROSINA during the entire duration of the \textit{Rosetta} mission between August 2014 and September 2016 are presented by \citet{Laeuter2020}. Based on a comparison of production rates, CH$_{3}$OH may be grouped with H$_{2}$O, O$_{2}$, H$_{2}$CO, and NH$_{3}$. Methanol production rate increases pre-perihelion, and then decreases post-perihelion at a slower rate than that of the increase upon approach of the Sun. However, H$_{2}$O does decrease much steeper than CH$_{3}$OH during the outward journey (for reasons that remain to be understood). The mission duration integrated production of CH$_{3}$OH (by number of molecules) is $0.5$ per cent of H$_{2}$O. The peak production rate of CH$_{3}$OH is also $\sim0.6$ per cent that of H$_{2}$O. There is no evidence to suggest that the methanol is more or less abundant in either of the hemispheres of 67P/C--G. There is also no evidence to suggest that the methanol isotopic ratios change over the course of the \textit{Rosetta} mission.


{\onecolumn
 \begin{center}
 \topcaption{Calculated D/H ratios in methanol for different assumptions based upon the average D-methanol/methanol and D$_{2}$-methanol/methanol ratios from the two studied data sets (early October and late October--December 2014), and the final range spanned by the ROSINA measurements of 67P/C--G. Statistical corrections for the location of D in the molecule and statistical error propagation in the ROSINA measurements are included here.}
 \label{tabl_ROSINA_DH}
 \tablefirsthead{\hline \multicolumn{1}{l}{ } & \multicolumn{1}{l}{\textbf{Assumption}} & \multicolumn{1}{l}{\textbf{D/H in CH$\mathbf{_{3}}$OH ($\mathbf{\%}$)}} \\ \hline}
 \begin{xtabular*}{\textwidth}{l@{\extracolsep{\fill}}lll}
 based on D-methanol/methanol         &                                                     & \T\\
                                      & all D-methanol is CH$_{3}$OD                        & $5.5\pm1.1$\\
                                      & all D-methanol is CH$_{2}$DOH                       & $1.8\pm0.2$\\
                                      & equally probable deuteration in OH-R and CH$_{3}$-R & $1.4\pm0.3$\B\\
 \hline
 based on D$_{2}$-methanol/methanol   &                                                     & \T\\
                                      & all D$_{2}$-methanol is CH$_{2}$DOD                 & $1.5\pm0.2$\\
                                      & all D$_{2}$-methanol is CHD$_{2}$OH                 & $1.5\pm0.2$\\
                                      & equally probable deuteration in OH-R and CH$_{3}$-R & $1.1\pm0.1$\B\\
 \hline
 based on D$_{2}$-methanol/D-methanol &                                                     & \T\\
                                      & equally probable deuteration in OH-R and CH$_{3}$-R & $0.83\pm0.12$\B\\
 \hline
 \textbf{Range}                       &                                                     & $\mathbf{\left[0.71, 6.6\right]}$\T\B\\
 \hline
 \end{xtabular*}
 \end{center}
\twocolumn}

\subsection{Methanol D/H ratio in 67P/C--G}
\label{DH_rat}

It is not possible to distinguish CH$_{3}$OD from CH$_{2}$DOH in the mass spectra from ROSINA at $m/z = 33$. The measured D-methanol/methanol ratio means that methanol deuteration fraction (i.~e., its D/H ratio) can be either $5.5\pm1.1$ or $1.8\pm0.2$ per cent, depending on if all the deuterium is in the hydroxyl group or the methyl group, respectively. This accounts for the statistical correction by a factor of $3$ when the D replaces one of the three identical hydrogens in CH$_{3}$-R. Likewise, it is not possible to distinguish CH$_{2}$DOD from CHD$_{2}$OH in the mass spectra from ROSINA at $m/z = 34$. The measured D$_{2}$-methanol/methanol ratio means that the D/H ratio is $1.5\pm0.2$ per cent, if all the deuterium is entirely in either CH$_{2}$DOD or CHD$_{2}$OH. This accounts for the statistical correction, which is the same in both of these cases. If both D atoms are in the methyl group, the statistical correction is given by:
\begin{equation}
\text{CHD}_{2}\text{OH}/\text{CH}_{3}\text{OH} = \frac{3!}{1! 2!} \times \left( \frac{\text{D}}{\text{H}} \right)^{2} = 3 \left( \frac{\text{D}}{\text{H}} \right)^{2}.
\end{equation}
If one D atom is in the methyl group and the other one is in the hydroxyl group, the statistical correction is given by:
\begin{equation}
\text{CH}_{2}\text{DOD}/\text{CH}_{3}\text{OH} = 3 \frac{\text{D}}{\text{H}} \times \frac{\text{D}}{\text{H}} = 3 \left( \frac{\text{D}}{\text{H}} \right)^{2}.
\end{equation}
Therefore, the D/H ratio in di-deuterated methanol is independent of the location of the two D atoms (for additional details see Appendix~B in \citealt{Manigand2019}). These D/H ratios correspond to extreme scenarios of mono- and di-deuterated methanol being dominated by just one specific variant.


Alternatively, it can be assumed that D can replace H with the same probability in the methyl and hydroxyl groups, then:
\begin{equation}
\left( \text{CH}_{3}\text{OD} + \text{CH}_{2}\text{DOH} \right) / \text{CH}_{3}\text{OH} = \frac{\text{D}}{\text{H}} + 3 \frac{\text{D}}{\text{H}} = 4 \frac{\text{D}}{\text{H}},
\end{equation}
\begin{equation}
\left( \text{CH}_{2}\text{DOD} + \text{CHD}_{2}\text{OH} \right) / \text{CH}_{3}\text{OH} = 3 \left( \frac{\text{D}}{\text{H}} \right)^{2} + 3 \left( \frac{\text{D}}{\text{H}} \right)^{2} = 6 \left( \frac{\text{D}}{\text{H}} \right)^{2}.
\end{equation}
Based on the measured D-methanol/methanol ratio, this yields a D/H of $1.4\pm0.3$ per cent. Based on the measured D$_{2}$-methanol/methanol ratio, this yields a D/H of $1.1\pm0.1$. The ratio of the measurements of D$_{2}$-methanol/methanol and D-methanol/methanol, i.e., the D$_{2}$-methanol/D-methanol ratio ($=\frac{3}{2}\frac{\text{D}}{\text{H}}$), yields a D/H of $0.83\pm0.1$ per cent.

Based on the available data, it is not possible to judge which of the above approaches to the calculation of the D/H ratio is more reliable. It cannot be, a priori, assumed that just one variant of a specific deuterated methanol isotopologues is present in the comet. Neither can it be safely assumed that deuteration occurs with the same probability in the two functional groups of methanol. The detection of di-deuterated methanol asserts that deuteration must occur in the methyl group, as both deuteriums cannot sit in the hydroxyl group. However, this does not necessarily impose where the deuterium is in mono-deuterated methanol, as the chemical pathways towards D$_{1}$- and D$_{2}$-methanol differ (Section~\ref{meth_chem}, Appendix~\ref{meth_Dchem}). The six possible values of D/H are summarized in Table~\ref{tabl_ROSINA_DH}. The errors on the D/H ratios are derived based on statistical error propagation of the statistical ($100\% / \sqrt{N}$, where $N$ is the number of ion counts on a certain $m/z$) and fit (none for normal methanol, $10\%$ for D$_{1}$-methanol, $15\%$ for D$_{2}$-methanol) errors, and then an inclusion of $11\%$ systematic error in the final step for normal methanol (in the cases when it was used for the derivation of the D/H). It is not possible to deduce a single D/H value based on the ROSINA data, rather only a range of $0.71-6.6$ per cent, which accounts for the location of D in the molecule and includes statistical error propagation in the ROSINA measurements. This range is wide primarily because of the various possible assumptions in the calculation of the D/H ratio.



\subsection{Mono-deuterated methanol from star-forming regions to comets}
\label{Dmeth}

\begin{figure*}
 \centering
  \includegraphics[width=0.8\textwidth,height=0.8\textheight,keepaspectratio]{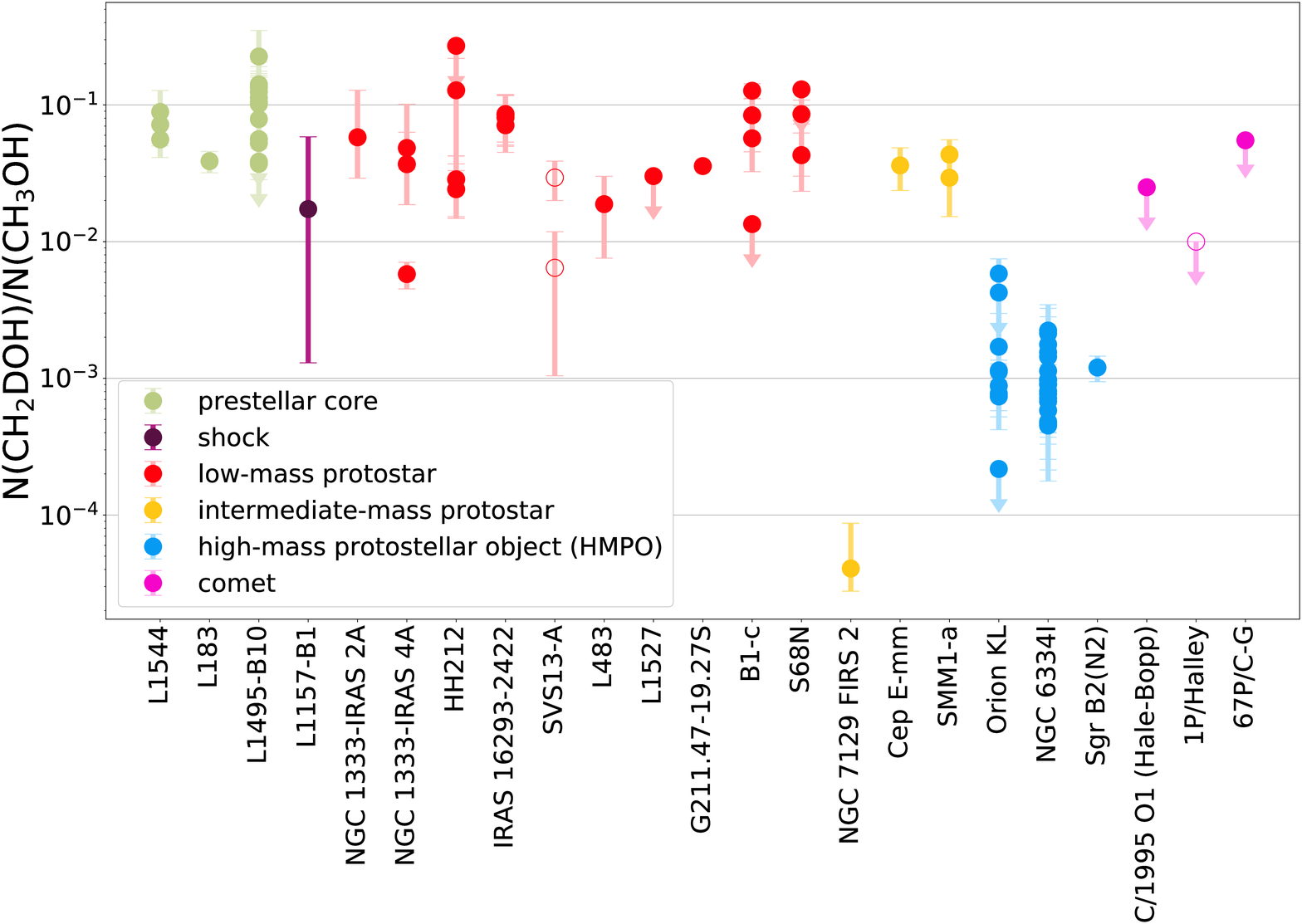}
 \caption{The ratio of CH$_{2}$DOH to CH$_{3}$OH towards star-forming regions and in comets, color-coded by the type of source. Single dish observations of low-mass protostars and high-mass star-forming regions (not prestellar cores) have been indicated with open circles, while interferometric observations are marked by filled circles. The mass spectrometry measurement for comet 1P/Halley is also indicated with an open circle due to the complication by the required complex ion modeling (Section~\ref{cavCom}). Full references are given in Table~\ref{tabl_Dmeth}. An exhaustive version is shown in Fig.~\ref{fig_CH2DOH_all}.}
 \label{fig_CH2DOH}
\end{figure*}

\begin{figure*}
 \centering
  \includegraphics[width=0.8\textwidth,height=0.8\textheight,keepaspectratio]{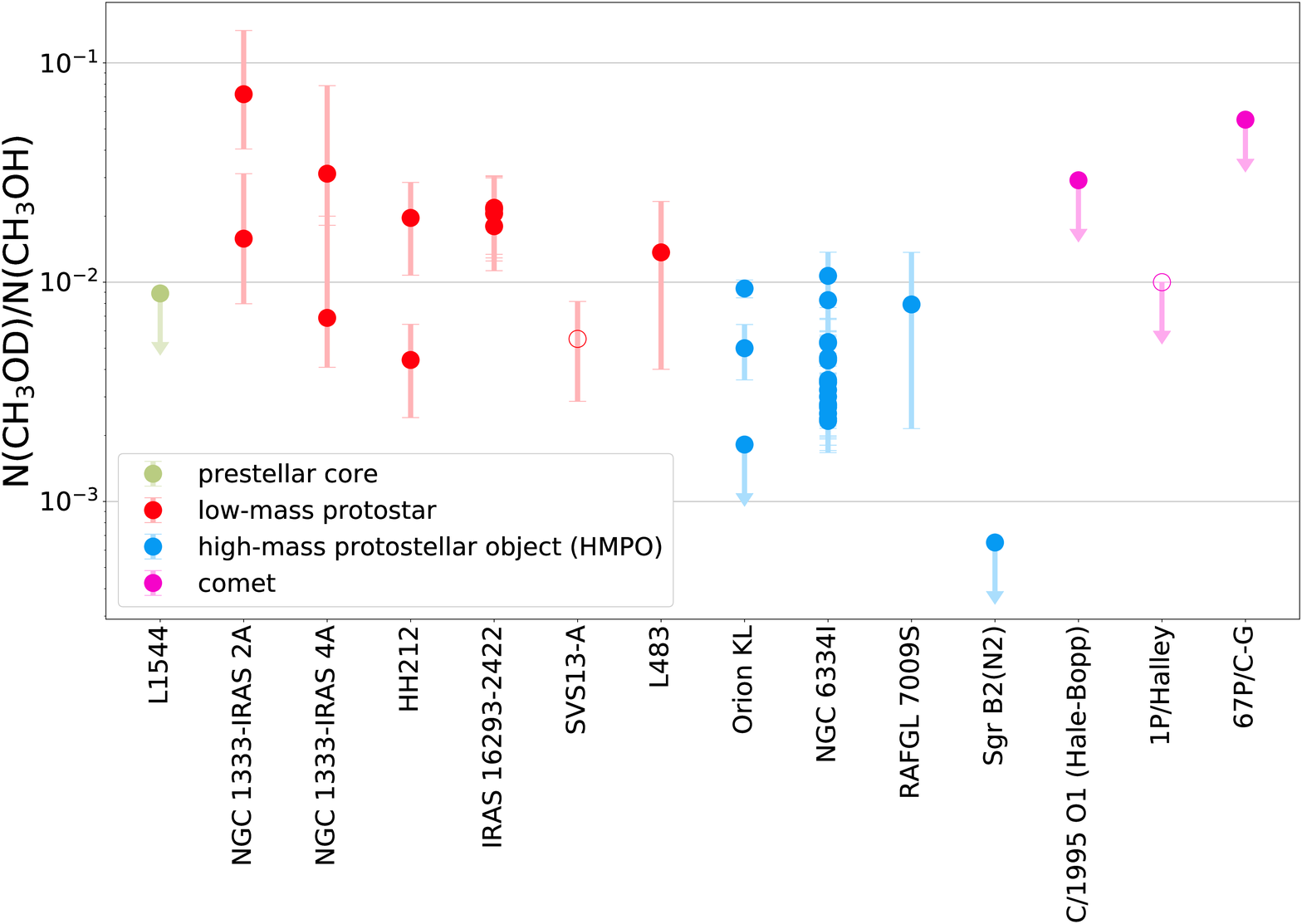}
 \caption{The ratio of CH$_{3}$OD to CH$_{3}$OH towards star-forming regions and in comets, color-coded by the type of source. Single dish observations of low-mass protostars and high-mass star-forming regions (not prestellar cores) have been indicated with open circles, while interferometric observations are marked by filled circles. The mass spectrometry measurement for comet 1P/Halley is also indicated with an open circle due to the complication by the required complex ion modeling (Section~\ref{cavCom}). Full references are given in Table~\ref{tabl_Dmeth}. An exhaustive version is shown in Fig.~\ref{fig_CH3OD_all}.}
 \label{fig_CH3OD}
\end{figure*}

\begin{figure*}
 \centering
  \includegraphics[width=0.8\textwidth,height=0.8\textheight,keepaspectratio]{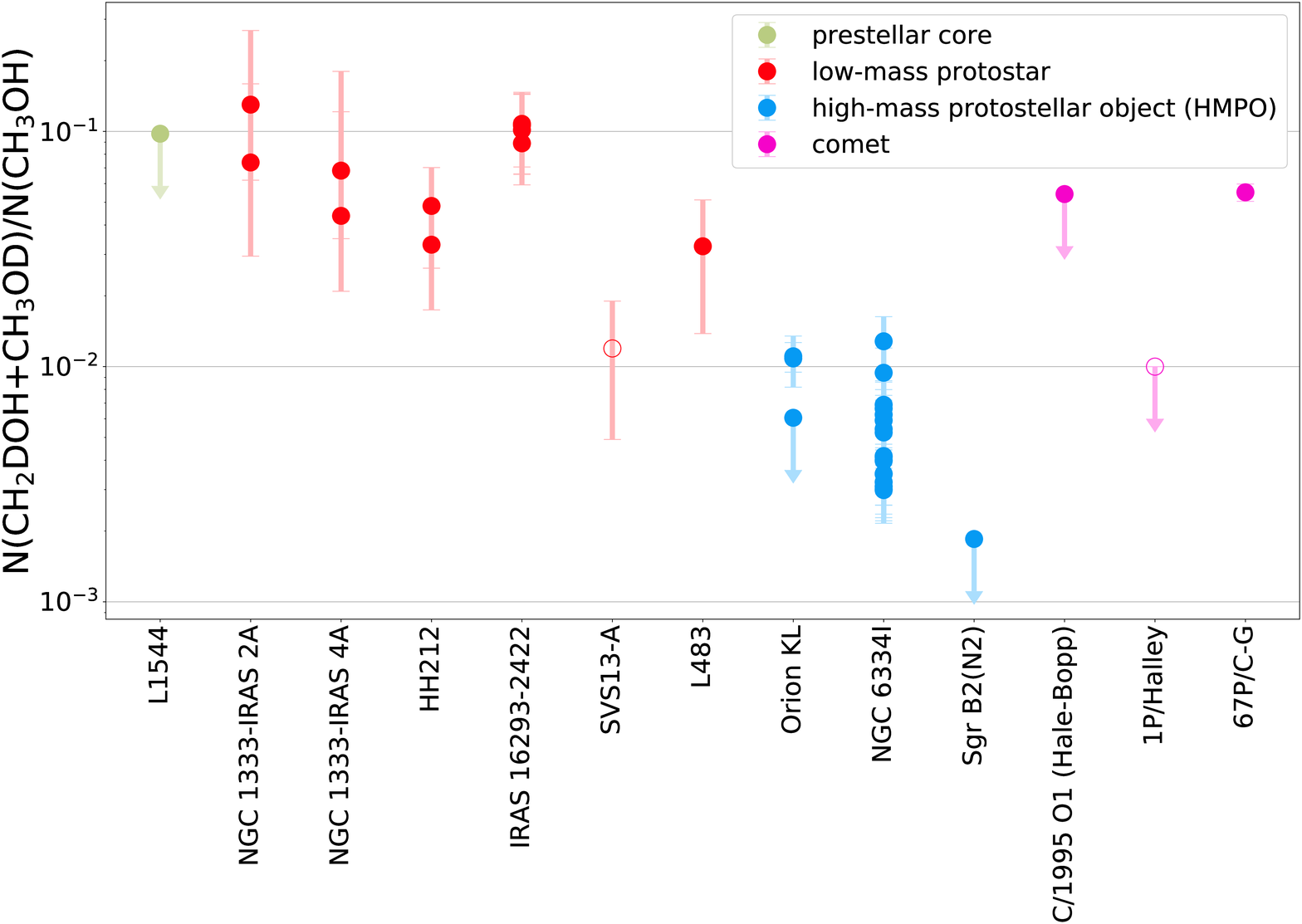}
 \caption{The ratio of CH$_{2}$DOH+CH$_{3}$OD to CH$_{3}$OH towards star-forming regions and in comets, color-coded by the type of source. Single dish observations of low-mass protostars and high-mass star-forming regions (not prestellar cores) have been indicated with open circles, while interferometric observations are marked by filled circles. The mass spectrometry measurement for comet 1P/Halley is also indicated with an open circle due to the complication by the required complex ion modeling (Section~\ref{cavCom}). Full references are given in Table~\ref{tabl_Dmeth}. An exhaustive version is shown in Fig.~\ref{fig_Dmethtot_all}.}
 \label{fig_Dmethtot}
\end{figure*}

In order to compare this first-time measurement of methanol deuteration in comets to that in distant star-forming regions to explore any possible trends, an extensive literature search has been carried out compiling all available observations and estimates of deuterated methanol relative to CH$_{3}$OH. Figs.~\ref{fig_CH2DOH}--\ref{fig_Dmethtot} show the ratio of CH$_{2}$DOH, CH$_{3}$OD, and CH$_{2}$DOH+CH$_{3}$OD relative to CH$_{3}$OH in prestellar and protostellar phases of low- and high-mass star formation, and in comets. Derivations of column densities of interstellar methanol and its isotopologues come with several caveats (discussed in detail in Appendix~\ref{caveats}). The smallest uncertainties stem from spectroscopy ($<10$ per cent for $T<150$~K and $<20$ per cent for $T<300$~K for all methanol variants as long as the rigid rotor approximation is not adopted; priv. comm. H.~S.~P.~M{\"u}ller) and from the assumption of LTE (less than a factor of $2$). However, \citet{Taquet2019} showed that if the rigid rotor approximation, that is to be avoided for methanol, is made for the partition function of CH$_{3}$OD, differences as large as a factor of $5$ in its column density can be incurred at $T=150$~K. The largest uncertainties in derived column densities originate from optical depth effects of the observed methanol gas and dust along the line of sight. If not treated with care, column densities can be easily underestimated by as much as an order of magnitude (most problematic near protostars for the case of normal methanol). It has not been entirely excluded that a molecule and its isotopologues may have different spatial distributions. Beam dilution effects (most drastic for single dish observations) will only cancel out if the spatial distributions are the same. More observations at higher spatial resolutions are needed to fully characterize the spatial distribution of methanol and its isotopologues. Finally, it should be mentioned that sources external to a system being observed may be influencing it in ways that are difficult to quantify. For example, the compact ridge in Orion~KL is subject to external heating sources \citep{Blake1987, Wang2011, Neill2013b}. Innate region to region variations may also lead to different column density ratios seen between low-mass protostars \citep{Bianchi2017b}. For these reasons, only the most-reliable interstellar methanol observations have been shown in Figs.~\ref{fig_CH2DOH}--\ref{fig_Dmethtot}. An exhaustive comparison is shown in Figs.~\ref{fig_CH2DOH_all}--\ref{fig_Dmethtot_all}. References are tabulated in Table~\ref{tabl_Dmeth} with additional details provided in Appendix~\ref{quantities}.

In the case of CH$_{2}$DOH, Fig.~\ref{fig_CH2DOH} shows that the birth places of low-mass protostars, i.~e., prestellar cores, tend to have a CH$_{2}$DOH/CH$_{3}$OH ratio in the $\sim10^{-1} - 10^{-2}$ range. Ratios predominantly in this range are also observed for low-mass protostars. High-mass protostellar objects (HMPOs) span a wider range of $\sim10^{-2} - 10^{-4}$ with values clustering around $\sim10^{-3}$. The data point corresponding to the protostellar shock L1157-B1 \citep{Codella2012} in a low-mass star-forming region agrees with ratios measured in prestellar cores and in later low-mass protostellar stages within its large error bars, but making its exact grouping with either of the two difficult. The measurement towards intermediate-mass protostar NGC~7129~FIRS~2 \citep{Fuente2014} differs by three orders of magnitude from that towards the other two intermediate-mass protostars, Cep~E-mm \citep{Ospina-Zamudio2018} and SMM1-a (Ligterink et al. 2020d in prep.). The former shows a closer agreement with high-mass sources, while the latter two with low-mass star-forming regions. If D-methanol in 67P/C--G would be dominated by CH$_{2}$DOH with only a smaller contribution from CH$_{3}$OD, then the 67P/C--G data point would not be much lower in Fig.~\ref{fig_CH2DOH}. The cometary CH$_{2}$DOH/CH$_{3}$OH ratio appears to agree well with that measured in prestellar cores and low-mass protostellar regions.

Whilst the CH$_{2}$DOH/CH$_{3}$OH ratio spans roughly three orders of magnitude, the CH$_{3}$OD/CH$_{3}$OH ratio spans roughly only two (Fig.~\ref{fig_CH3OD}). The prestellar core, low-mass protostellar, HMPO, and cometary ratios are all lying around the $\sim10^{-2}$ value. The ratios measured in HMPOs do tend to be clustered at values that are lower than those measured in low-mass protostars by a factor of a few (but not exceeding a factor of $10$). The variations in the CH$_{3}$OD/CH$_{3}$OH ratio appear to be smaller than in the CH$_{2}$DOH/CH$_{3}$OH ratio across phases and mass regimes of star formation.

When the sum of CH$_{2}$DOH and CH$_{3}$OD relative to CH$_{3}$OH ratio is investigated, the sources with at least an estimate of CH$_{3}$OD adopt the larger spread of CH$_{2}$DOH/CH$_{3}$OH values. The (CH$_{2}$DOH+CH$_{3}$OD)/CH$_{3}$OH ratio of HMPOs is roughly a factor of $10$ lower than that in low-mass protostars. The tightly constrained measurement of 67P/C--G agrees with the observed ratios in L483 (of \citealt{Agundez2019}), NGC~1333-IRAS~2A (of \citealt{Taquet2019}) and~-IRAS~4A (measured for the binary by \citealt{Taquet2019} and for A2 specifically by \citet{Sahu2019}, see Appendix~\ref{caveats}), HH212 (of \citealt{Taquet2019}), and IRAS~16293-2422~A (of \citealt{Manigand2020a} when using the $^{18}$O-isotopologue to estimate the CH$_{3}$OH column density). The ratio for IRAS~16293-2422~B is a factor of $\sim1.6$ higher than in 67P/C--G (of \citealt{Jorgensen2018} also when using the $^{18}$O-isotopologue to estimate the CH$_{3}$OH column density). The less-constrained ratio for SVS13-A is a factor of $\sim4.6$ lower than in 67P/C--G\footnote{This measurement towards SVS13-A stems from single dish IRAM~30m observations with the intrinsic caveats of this type of observations and the undesirable use of the rigid rotor approximation for the partition function of CH$_{3}$OD (Appendix~\ref{caveats}). The data stem from an unbiased spectral survey, which allowed the methanol emission to be separated into two components. CH$_{3}$OD has only been detected for the ``low-temperature $3\arcsec$-component'', and thus the ratio has been computed solely for this component. Most importantly, the optically thin $^{13}$CH$_{3}$OH was used to derive the CH$_{3}$OH column density in these observations. Finally, SVS13-A is one of the two sources with di-deuterated methanol detections (Section~\ref{DDmeth}), and consequently, it is displayed in Figs.~\ref{fig_CH2DOH}--\ref{fig_Dmethtot}, alongside the more-reliable interferometric data.}.

\subsubsection{Caveats of cometary measurements}
\label{cavCom}

Minor isotopologues such as deuterated methanol are rarely searched for with remote observations of comets due to their low abundance, and the anticipated long integration times. Thus far, only two upper limits have been reported in the literature. One stems from remote observations of comet C/1995 O1 (Hale--Bopp) carried out with the Caltech Submillimeter Observatory (CSO) and IRAM~30m telescopes \citep{Crovisier2004}. The caveats of interstellar observations described in Appendix~\ref{caveats} apply to remote cometary observations in analogous ways. Non-interferometric observations struggle to constrain the spatial distribution of species in a coma, which will affect abundances derived for secondary species (those that are produced by chemical processes within the coma) or species originating from distributed sources (i.~e., dust grains that are lifted into the coma). For methanol, this is expected to be of minor importance, and a Haser distribution \citep{Haser1957} may be assumed. Typically, as in \citet{Crovisier2004}, coma molecules are assumed to be in LTE with a constant excitation temperature in the coma. As gas and dust densities are relatively low in cometary coma, gas and dust optical depths are not a grave concern. The spectroscopic uncertainties discussed in Appendix~\ref{caveats} apply fully to cometary remote observations in the same way as to interstellar objects.

Mass spectrometry measurements at small cometocentric distances, such as those associated with the data presented in this work from \textit{Rosetta}--ROSINA, shed light on the frozen-out volatiles in the interior of a cometary nucleus before any additional chemical processing can take place in the coma. However, it remains difficult to piece together the spatial distribution of desorbing gas as the measurements are being carried out at one specific location above a certain part of the comet (e.~g., \citet{Laeuter2020}). Mass spectrometry eliminates the uncertainties stemming from excitation and spectroscopy. However, it does come with its intrinsic difficulty of unique assignments of mass peaks to parent and daughter species, as well as its inability to distinguish isotopologues (such as CH$_{3}$OD and CH$_{2}$DOH). A careful analysis in conjunction with a high mass resolution can facilitate a firm, well-constrained assignment of the mass peaks, as in the case of the ROSINA data presented in this work. Methanol data from the Neutral Mass Spectrometer (NMS) on the \textit{Giotto} spacecraft for comet 1P/Halley presented by \citet{Eberhardt1994} was obtained by analyzing ions. The coma of 1P/Halley was much more dense than that of 67P/C--G with many chemical reactions taking place, especially proton transfer giving rise to CH$_{3}$OH$_{2}^{+}$. Since molecular oxygen does not protonate readily, it is likely that $m/z = 33$ is dominated by CH$_{3}$OH$_{2}^{+}$ and $m/z = 34$ by protonated D-methanol. However, the interpretation of these results requires a chemical model and carries much larger intrinsic uncertainties \citep{Rubin2015b} than those from ROSINA, which probed the neutral species directly (e.~g.,~Fig.~\ref{fig_ROSINA_32}).

\subsection{Di-deuterated methanol from star-forming regions to comets}
\label{DDmeth}

There are only two detections of CHD$_{2}$OH in the ISM, namely towards low-mass protostars IRAS~16293-2422 and SVS13-A with column density ratios of $0.17$ and $0.092-0.00074$ relative to normal methanol, respectively \citep{Parise2002, Bianchi2017a}. CH$_{2}$DOD has never been detected in the ISM; however, spectroscopic data in the mm-submm wavelength range that would enable its search are not available. The combined measurement of CH$_{2}$DOD and CHD$_{2}$OH by ROSINA in comet 67P/C--G at an abundance of $0.00069\pm0.00014$ relative to normal methanol is in agreement within errors with part of the range derived for SVS13-A. For SVS13-A, only single dish observations with the IRAM~30m telescope are available in the literature for both, mono- and di-deuterated methanol. A measurement of CH$_{3}$OD in SVS13-A is only available for the ``low-temperature $3\arcsec$-component'' of \citet{Bianchi2017a}, while CH$_{2}$DOH has been measured in the ``low-temperature $3\arcsec$-component'' and the ``high-temperature $0.3\arcsec$-component''. The (CH$_{2}$DOH+CH$_{3}$OD)/CH$_{3}$OH ratio of the ``low-temperature $3\arcsec$-component'' in SVS13-A is lower than the cometary value (Fig.~\ref{fig_Dmethtot}). Such a component-wise analysis was not possible based on the single dish data for IRAS~16293-2422 in \citet{Parise2002}, making it impossible to pinpoint the origins of the different di-deuterated methanol abundances without dedicated interferometric observations. Most importantly, the optically thin $^{13}$CH$_{3}$OH was used to derive the CH$_{3}$OH column density in SVS13-A observations in \citet{Bianchi2017a}, contrary to the IRAS~16293-2422 observations of \citet{Parise2002}. The uncertainties in the column densities of CHD$_{2}$OH in star-forming regions are grossly exacerbated in comparison to those associated with mono-deuterated methanol (Appendix~\ref{caveats}) most drastically in regards to the rigid rotor approximation made for the partition function and the lack of readily available precise spectroscopy. A more thorough comparison between existing ISM and cometary quantities of di-deuterated methanol is thus currently not feasible.

\section{Discussion}
\label{discussion}

\subsection{Formation and deuteration of methanol}
\label{meth_chem}

In the sequence of events that form stars and protoplanetary discs, methanol is first formed in the earliest prestellar phase. It is observed in prestellar cores as a gas (e.~g., in L1544, \citealt{Bizzocchi2014, Chacon-Tanarro2019, Lattanzi2020}, in L183, \citealt{Lattanzi2020}, and in L1595, \citealt{ScibelliShirley2020}), and also as an ice in prestellar and starless cores \citep{Boogert2011, Boogert2015}. Laboratory experiments have verified that methanol forms on grain surfaces via sequential hydrogenation of CO \citep{WatanabeKouchi2002b, Fuchs2009} with efficient gas-phase pathways being ruled out \citep{Geppert2006}. Under the physical conditions of cores, the chemistry of solid methanol should be dominated by H atom additions as there are no internal protostellar UV sources and shielding from external UV sources is high. Experiments have shown that CH$_{3}$OH does not undergo efficient reactive desorption during hydrogenation of CO (only on the order of a few per cent, \citealt{Chuang2018a}), which is drastically reduced for a surface of amorphous water ice \citep{Hidaka2008}. Cosmic rays are the sole source of UV photons in cores, because they can impact H$_{2}$ molecules, which then de-excite via a fluorescence cascade \citep{PrasadTarafdar1983}. Consequently, gaseous CH$_{3}$OH in cores is likely a testament of the small, but prolonged, desorbing influence of cosmic rays (either via spot heating, \citealt{Ivlev2015a}, or photodesorption by CR-induced UV photons). Although it has been shown in laboratory studies that methanol does not photodesorb intact, the released photofragments may pave the way to methanol reformation in the gas phase \citep{Bertin2016}. The importance of cosmic rays for the chemistry in cores is also supported by dedicated theoretical works (e.~g., \citealt{Shingledecker2017, Shingledecker2018b}).

For hydrogenation of solid CO to occur, H atoms must diffuse across the dust grain surface via thermal hopping and quantum tunneling to meet a CO molecule to form HCO while overcoming the reaction barrier via quantum tunneling. Experiments have shown that H (and D) atom diffusion is dominated by thermal hopping on amorphous water ice and on pure CO ice \citep{Hama2012, Kimura2018}. The formation of H$_{2}$CO by addition of H to the HCO radical is barrierless. The subsequent addition of H to H$_{2}$CO may theoretically form either the methoxy (CH$_{3}$O) or the hydroxymethyl (CH$_{2}$OH) radicals. The formation of CH$_{2}$OH is more exothermic than that of CH$_{3}$O; however, it has a higher activation barrier \citep{Woon2002, Osamura2004}. Thus, CH$_{2}$OH formation is less likely than that of CH$_{3}$O, although not entirely excluded \citep{Chuang2016}. The final H addition to either of these radicals to form methanol is again barrierless. Dust temperature, CO/H$_{2}$O ice purity, the relative ratio of CO:H$_{2}$CO:CH$_{3}$OH, the abundance of atomic H, and simultaneous UV-photolysis change the efficacy of hydrogenation reactions \citep{Watanabe2004, Watanabe2006, Chuang2016, Chuang2017}.

In star-forming regions, D atoms are available alongside H atoms for grain-surface chemistry. Methanol deuteration reaction schemes depend on the order in which deuteration is to occur, and are discussed in light of the suite of executed laboratory work in Appendix~\ref{meth_Dchem}. In summary, CH$_{2}$DOH is thought to be formed from CO when both H and D atoms are available for its synthesis: CH$_{3}$OH being formed first via hydrogenations, and then subsequent deuteration in the methyl group occurring through H-D substitution reactions (H abstraction followed by D addition, \citealt{Nagaoka2005}). CHD$_{2}$OH is formed along this forward synthesis pathway as well, via H-D substitution reactions in formaldehyde and subsequent hydrogenation of D$_{2}$CO \citep{Hidaka2009}. On the long time-scales of cores and star formation, CH$_{3}$OD may form starting from non-deuterated formaldehyde and the CH$_{3}$O radical (that is preferentially produced by H$_{2}$CO+H), but likely at a very low rate as H addition proceeds more efficiently than that of D. It is thought that this is the sole time that the CH$_{3}$O radical is present in the solid phase \citep{Nagaoka2007, GoumansKaestner2011}. Alternatively, CH$_{3}$OD may form upon an isotope exchange reaction between non-deuterated methanol and deuterated water or deuterated ammonia \citep{Kawanowa2004}. Laboratory experiments have shown that the hydroxyl group of deuterated methanol undergoes H-D exchange reactions with non-deuterated water due to its ability to hydrogen bond unlike the methyl group \citep{Souda2003, Souda2004, Ratajczak2009, FaureM2015}. This could then also explain the formation of CH$_{2}$DOD, as the CH$_{2}$DO radical is unlikely to be present otherwise. Reaction schemes of the deuterated chemical network for methanol are visualized in fig.~$1$ of \citet{Hidaka2009} and fig.~$8$ of \citet{Chuang2016}, for example.

Once a protostar is born, the physical conditions change drastically in comparison to those during the prestellar stage. Internal UV irradiation starts to play a critical role in the grain-surface chemistry of methanol before its thermal desorption into the gas phase in regions that are warmer than $\sim100$~K. At lukewarm ($\sim40-60$~K) dust temperatures, associations of heavier radicals become efficient as their mobility on the grain surfaces increases (e.~g., \citealt{Watanabe2007, Oberg2009}). In this temperature regime, methanol formation is dominated by the association of CH$_{3}$ and OH on grain surfaces rather than hydrogenations. Potentially the availability and mobility of CH$_{3}$O for synthesis into CH$_{3}$OD is also enhanced; however, the residence time of deuterium atoms is much shorter at these temperatures. Deuteration of methanol is likely fully halted at temperatures above $\sim20$~K. Besides, the detections of CH$_{2}$DOH and CH$_{3}$OD at cold ($\sim10-20$~K) conditions imply that there must be a low-temperature, low-UV formation and deuteration pathways.

\subsection{Formation and deuteration of volatiles in 67P/C--G}
\label{chem_impl}

Comet 67P/C--G likely carries both variants of mono-deuterated methanol (CH$_{3}$OD, CH$_{2}$DOH), and both variants of di-deuterated methanol (CH$_{2}$DOD, CHD$_{2}$OH) at combined abundances of $5.5\pm0.46$ and $0.069\pm0.014$ per cent relative to CH$_{3}$OH, respectively (Section~\ref{mass_spec}). Methanol is available in the coma of 67P/C--G at an average level of $0.5$ per cent relative to water (\citealt{Laeuter2020}; Section~\ref{mass_spec}). The methanol formation and deuteration schemes are supported by a wealth of laboratory experiments (Section~\ref{meth_chem}, Appendix~\ref{meth_Dchem}). These schemes imply that the methanol found in comet 67P/C--G must have formed from CO in the presence of H and D atoms on grain surfaces (yielding CH$_{3}$OH, CH$_{2}$DOH, CHD$_{2}$OH). Possibly, this also occurred embedded in an ice containing deuterated water (yielding CH$_{3}$OD, CH$_{2}$DOD). Both mono- and di-deuterated water have been detected in 67P/C--G \citep{Altwegg2015, Altwegg2017a}. Although methanol and its isotopologues are minor in their overall abundance within the cometary coma, these species are a critical testament to the comet's cold formative past. Temperatures had to be low enough to sustain CO ice and to ensure long residence times of H and D atoms on grain surfaces for reactions to occur. The presence of di-deuterated methanol with at least one deuterium in the methyl group suggests that deuterated formaldehyde should also be present in the comet. Unfortunately, HDCO is not distinguishable from H$_{2}^{13}$CO in the CH$_{3}$O-dominated peak; and D$_{2}$CO is not distinguishable from H$_{2}$C$^{18}$O in the CH$_{3}$OH-dominated peak (Fig.~\ref{fig_ROSINA_32}).

The ratio of D$_{2}$-methanol/D-methanol to D-methanol/methanol in comet 67P/C--G $0.23\pm0.060$ based on data presented here (with statistical error propagation). Assuming equally probable deuteration in both functional groups, the statistically expected ratio is $0.375$ based on the equations in Section~\ref{DH_rat}. Thus, the measured ratio is a factor of $1.3-2.2$ lower than the statistically expected number. This is contrary to the case of water, because the measured ratio of D$_{2}$-water/D-water to D-water/water is $\sim17$, while the statistically expected value is $0.25$ \citep{Altwegg2017a}. So for water, the measured ratio is a factor of $68$ higher than the statistically expected number. The physicochemical models of \citet{Furuya2016, Furuya2017} explain this by the bulk of H$_{2}$O forming in molecular clouds prior to the formation of HDO and D$_{2}$O in prestellar cores. It was also postulated that D-methanol/methanol~$\sim$~D$_{2}$O/HDO~$>$~HDO/H$_{2}$O, as deuteration of methanol and HDO would be occurring at the same epoch on top of a thick bulk layer of H$_{2}$O ice. For 67P/C--G volatiles, D-methanol/methanol$~=0.055\pm0.0046$ is roughly a factor of $3$ higher than D$_{2}$O/HDO$~=0.0180\pm0.009$; while both of these values are an order of magnitude higher than HDO/H$_{2}$O$~=0.00105\pm0.00014$ \citep{Altwegg2015, Altwegg2017a}\footnote{Note that table~$1$ of \citet{Altwegg2017a} contains a typo in the D/H ratio of water. The correct values are given in the abstract and in subsequent publications. The water D/H ratio stated in \citet{Altwegg2015} and in table~$1$ of \citet{Altwegg2017a} accounts for the statistical correction by a factor of $2$.}. So the measurements point to the relation being D-methanol/methanol~$>$~D$_{2}$O/HDO~$\gg$~HDO/H$_{2}$O, i. e., partially supporting the postulation of \citet{Furuya2016, Furuya2017}. The ratios of HDS/H$_{2}$S and NH$_{2}$D/NH$_{3}$ of $0.0012\pm0.0003$ and $0.001$, respectively \citep{Altwegg2017a, Altwegg2019}, are in closer agreement with HDO/H$_{2}$O, suggesting that they undergo deuteration at the same time as water to mono-deuterated water. Unfortunately, the abundance of D$_{2}$S cannot be probed by ROSINA due to its peak being located directly under that of C$_{3}$ (fig.~$1$ of \citealt{Balsiger2015}). Physicochemical models of \citet{Furuya2017} show a small enhancement in the methanol D/H ratio as a result of further chemical processing during the collapse phase and in the cold midplane of the protoplanetary disc. If comet 67P/C--G contains significant fractions of volatiles from these evolutionary phases, then deuteration in the innate core that birthed our Solar System may have been slightly lower, and consequently, slightly warmer.

The formation of water ice in molecular clouds should occur around the same epoch as the formation of carbon monoxide gas, which would imply that the source of oxygen in H$_{2}$O is the same as that in CO. The oxygen in CO would then be transferred in the prestellar core stage via grain-surface chemistry into methanol and CO$_{2}$, suggesting that the oxygen isotopic ratios in H$_{2}$O, CH$_{3}$OH, and CO$_{2}$ should agree. This is supported by the data presented by \citet{Altwegg2020b, Schroeder2019b, Haessig2017}, where $^{16}$O/$^{18}$O are shown to agree within errors for these three molecules. The oxygen isotopic ratio in CO cannot be directly determined due to a mass overlap of C$^{18}$O with NO that cannot be separated at the resolving power of the DFMS. However, the carbon $^{12}$C/$^{13}$C isotopic ratio of CO can be determined and is shown to agree within errors with that of CO$_{2}$ and CH$_{3}$OH \citep{Rubin2017, Altwegg2020b}. This suggests that the source of carbon in CO$_{2}$ and CH$_{3}$OH matches that of CO. Consequently, supporting the grain-surface chemistry sequence once more. The oxygen and carbon isotopic ratios of formaldehyde have been shown to be a factor of $\sim2$ lower than those of H$_{2}$O, CH$_{3}$OH, CO$_{2}$, and CO \citet{Altwegg2020b}. This is evidence for the formation of H$_{2}$CO not only through grain-surface chemistry via the hydrogenation of CO, but also through gas-phase chemistry from carbon and oxygen reservoirs with lower $^{12}$C/$^{13}$C and $^{16}$O/$^{18}$O isotopic ratios. The gas-phase formation of H$_{2}$CO at low temperatures is thought to proceed predominantly through CH$_{3}$ + O $\longrightarrow$ H$_{2}$CO + H \citep{vanderTak2000, FockenbergPreses2002, Atkinson2006, vanderMarel2014}. One possible interpretation of the ROSINA volatile oxygen and carbon isotopic ratios could be that there are two reservoirs of carbon and oxygen in prestellar cores: (1) the grain-surface reservoir traced by H$_{2}$O, CH$_{3}$OH, CO$_{2}$, and CO, which matches the bulk gaseous reservoir of the molecular clouds; and (2) the gaseous reservoir traced in part by H$_{2}$CO, which is poor in the rare $^{13}$C and $^{18}$O isotopes. Deuteration of formaldehyde would only proceed for grain surface-formed H$_{2}$CO. If the formaldehyde deuteration fraction could be determined, then the D$_{2}$CO/HDCO ratio would be expected to be in close agreement with D-methanol/methanol and D$_{2}$O/HDO, while the HDCO/H$_{2}$CO ratio would be much lower.

\subsection{Volatiles of 67P/C--G as tracers of our Solar System's past}
\label{SS_impl}

In light of the relative deuteration fractions of methanol, water, hydrogen sulfide, and ammonia, and the carbon and oxygen isotopic ratios (Section~\ref{chem_impl}), the volatiles in comet 67P/C--G appear to be consistent with the molecular cloud to prestellar core to protostar evolutionary sequence of star formation: H$_{2}$O, CO, CO$_{2}$, H$_{2}$S, NH$_{3}$ form first in clouds as gases; HDO, HDS, NH$_{2}$D, CH$_{3}$OH are made via grain-surface chemistry in cores; and close to the onset of collapse, on the longest time-scales, D$_{2}$O and deuterated methanol isotopologues appear. Current robust understanding of the methanol chemical network under physical conditions of cores supports that the deuterated methanol isotopologues in 67P/C--G must be made when CO, H atoms, D atoms, and potentially deuterated water are available as solids (Section~\ref{meth_chem}). This is the case in environments that are cold enough to sustain volatile CO as an ice (temperatures $<20$~K). Such environments are also likely cold enough to enable efficient chemistry with H and D atoms when their residence times on the grains are long enough (optimised at $\sim15$~K, \citealt{Cuppen2009}). The methanol D/H ratio in comet 67P/C--G is in the $0.71-6.6$ per cent range (Section~\ref{DH_rat}, accounting for the location of D in the molecule and including statistical error propagation in the ROSINA measurements). This value is $2-3$ orders of magnitude higher than the elemental D/H ratio of the local ISM at just $\left( 2.0 \pm 0.1 \right) \times 10^{-5}$ (Section~\ref{intro}). Such special chemical circumstances can only be attained at specific physical conditions such as low temperatures. In comparison to the physicochemical model output of \citet{Taquet2012, Taquet2013, Taquet2014} presented in fig.~$8$ of \citet{Bogelund2018}, the methanol D/H ratio of 67P/C--G is consistent with temperatures below $\sim25$~K. This further supports that cometary methanol was formed in the dark, cold core that birthed our Solar System.

Modeling of thermophysics, hydrostatics, orbit evolution, and collision physics suggests that 67P/C--G is a primordial rubble pile \citep{Davidsson2016}. Its persistent exposure to temperatures below $20$~K is supported by the detection of not only CO, but also other hypervolatiles such as O$_{2}$ \citep{Bieler2015b, Fougere2016b, Gasc2017, Keeney2017, Noonan2018, Keeney2019a, Hoang2019, Combi2020, Laeuter2020}, N$_{2}$ \citep{Rubin2015a}, CH$_{4}$ \citep{LeRoy2015, Schuhmann2019}, and Ar \citep{Balsiger2015} with low binding energies \citep{Ayotte2001, Collings2004, Bar-Nun2007}. Some works argue that the only way that O$_{2}$ can be produced at the abundance level $\sim4$ per cent relative to water in 67P/C--G and show such a strong correlation with water is for it to be formed in a prestellar core that has a slightly elevated temperature of $\sim15-25$~K (in contrast to the typical core temperature of $10$~K) through a combination of gas-phase and grain-surface processes \citep{Taquet2016b, Taquet2018, Eistrup2019a}. However, other models claim this to not necessarily be the case \citep{Garrod2019, Rawlings2019}. Most of the alternative mechanisms summarized in \citet{Luspay-Kuti2018} have been ruled out (e.~g., \citealt{Altwegg2020b}). Potentially, the phase and mobility of the oxygen atom is a critical parameter for the O$_{2}$ chemical network \citep{vanDishoeck2014b}. The N$_{2}$/CO ratio suggests a lack of N$_{2}$ in 67P/C--G, which could also be interpreted as a result of a slightly elevated core temperature \citep{Rubin2015a} of $\sim20$~K (based on bulk abundances in \citealt{Rubin2019a}). On the other hand, it has been recently found that much of nitrogen could be hidden in the form of ammonium salts \citep{Altwegg2020a}, suggesting a revision of the current nitrogen chemical network. Finally, \citet{Calmonte2016} claimed to have recovered the undepleted molecular cloud sulphur elemental budget, pinpointing clouds as the source of cometary sulphur. The evidence for 67P/C--G being a relic our Solar System's cold past is overwhelming (as argued by many other publications, e.~g., \citealt{Alexander2018, Rubin2019b}); however, the exact temperature regime of the innate core cannot yet be claimed conclusively. The lower limit on the temperature of the birth core should correspond to the binding energy of neon, as this second-most volatile (after helium) noble gas has not been detected in 67P/C--G \citep{Rubin2018}.


\section{Conclusions}
\label{conclusions}

The ROSINA instrument aboard the ESA \textit{Rosetta} mission to JFC 67P/C--G detected mono- and di-deuterated methanol for the first time in a cometary coma. CH$_{3}$OH is present on average at $0.5$ per cent relative to H$_{2}$O \citep{Laeuter2020}, while D-methanol and D$_{2}$-methanol are measured to be at an abundance of $5.5\pm0.46$ and $0.069\pm0.014$ per cent relative to normal methanol. The data suggest that comet 67P/C--G likely carries both variants of mono-deuterated methanol (CH$_{2}$DOH and CH$_{3}$OD) and both variants of di-deuterated methanol (CH$_{2}$DOD, CHD$_{2}$OH), although it is not possible to identify the individual isotopologues directly from the mass spectra. A methanol deuteration fraction (D/H ratio) in the $0.71-6.6$ per cent range is spanned by the ROSINA data on mono- and di-deuterated methanol, accounting for statistical corrections for the location of D in the molecule and including statistical error propagation in the ROSINA measurements. This value is $2-3$ orders of magnitude higher than the elemental abundance of D relative to H of the local ISM.

Deuterated methanol is one of the most robust windows astrochemists have on the individual chemical reactions forming D-bearing volatiles due to a wealth of dedicated laboratory experiments and theoretical calculations (e.~g., \citealt{WatanabeKouchi2002b, Osamura2004, Nagaoka2005, Fuchs2009, Hidaka2009, Chuang2016, Chuang2017}). This paper suggests that the CH$_{2}$DOH in comet 67P/C--G stems from the hydrogenation chain of CO to CH$_{3}$OH, followed by H-D substitution reactions in the methyl functional group. Deuterium atoms were likely available simultaneously, consequently also forming CHD$_{2}$OH via chemical reactions involving deuterated formaldehyde. In this scenario, CH$_{3}$OD and CH$_{2}$DOD would form via H-D exchange reactions in the hydroxy functional group, if the cometary methanol is formed in or on top of an ice of deuterated water.

The D/H ratios, as well as the oxygen and carbon isotopic ratios, in methanol and other volatiles of 67P/C--G point towards a sequence of formation for the comet's molecules: H$_{2}$O, CO, CO$_{2}$, H$_{2}$S, NH$_{3}$ first in clouds as gases; HDO, HDS, NH$_{2}$D, CH$_{3}$OH second in cores as ices; and D$_{2}$O and deuterated methanol isotopologues last. This sequence is fully consistent with the evolutionary scenario of star-forming regions and is partially supported by the physicochemical models of \citealt{Furuya2016, Furuya2017}). Methanol and its deuterated isotopologues in comet 67P/C--G must have formed in the innate prestellar core that would go on to birth our Solar System at a time when it was at a temperature of $10-20$~K. Beyond the physicochemical arguments, this is also supported from the observational perspective. The tightly constrained ROSINA D-methanol/methanol ratio of comet 67P/C--G agrees more closely with those measured in prestellar cores and low-mass protostellar regions, specifically L483, NGC~1333-IRAS~2A and -IRAS~4A, HH212, and IRAS~16293-2422~A (meanwhile, the ratio in IRAS~16293-2422~B is a factor of $\sim1.6$ higher than in 67P/C--G). The cometary D$_{2}$-methanol/methanol ratio shows a tentative overlap with the lower end of the sparse ISM estimates (stemming from low-mass protostars).

Methanol is a pivotal precursor to complex organic molecules, and could be a source of D for such species \citep{Oba2016c, Oba2019b}. Since the donation of D differs depending on whether CH$_{2}$DOH or CH$_{3}$OD is the source \citep{Oba2017}, this could be a potential window on the exact synthesis of complex organic molecules in star-forming regions. As more observations at high sensitivity become available in the future, deuteration of complex organic molecules can be explored in light of deuterated methanol. The presented findings should also be used to stimulate deep characterizations of other cometary coma and their minor constituents such as deuterated methanol.

\section{Acknowledgements}
\label{acknowledgements}

This work is supported by the Swiss National Science Foundation (SNSF) Ambizione grant 180079, the Center for Space and Habitability (CSH) Fellowship, and the IAU Gruber Foundation Fellowship. MR acknowledges the State of Bern and the Swiss National Science Foundation (SNSF) under grant 200020\_182418. JDK acknowledges support by the Belgian Science Policy Office via PRODEX/ROSINA PEA 90020. Research at Southwest Research Institute was funded by NASA Grant 80NSSC19K1306. Work at UoM was supported by contracts JPL 1266313 and JPL 1266314 from the NASA US \textit{Rosetta} Project.

ROSINA would not have produced such outstanding results without the work of the many engineers, technicians, and scientists involved in the mission, in the \textit{Rosetta} spacecraft team, and in the ROSINA instrument team over the last 25 years, whose contributions are gratefully acknowledged. \textit{Rosetta} is an ESA mission with contributions from its member states and NASA. We acknowledge herewith the work of the whole ESA \textit{Rosetta} team.

The authors would like to acknowledge the contributions to this work of S{\'e}bastien Manigand with regards to the derivation of D/H ratios from multiply deuterated molecules; Holger~S.~P. M{\"u}ller with regards to the spectroscopy of methanol and its isotopologues; Niels F.~W. Ligterink and Vianney Taquet in connection to discussions about laboratory and theoretical aspects of methanol and deuterated methanol chemistry, respectively; and Florian Reinhard with regards to discussions about molecular D/H ratios. This work benefited from discussions held with the international team \#461 ``Provenances of our Solar System’s Relics'' (team leaders Maria~N. Drozdovskaya and Cyrielle Opitom) at the International Space Science Institute, Bern, Switzerland.

\section{Data availability}

The ROSINA data underlying this article are available from the ESA \textit{Rosetta} data archive at \url{https://www.cosmos.esa.int/web/psa/rosetta} and its mirror site at NASA Small Bodies Node \url{https://pds-smallbodies.astro.umd.edu/data_sb/missions/rosetta/index.shtml}. All further data processing is as described in the main body and the appendices of the article.

\clearpage
\bibliographystyle{mn2e}
\bibliography{mybib} 

\clearpage
\newpage
\appendix

\section{67P/C--G: supplementary ROSINA mass spectra and data}
\label{mass_spec_Oct14}

\begin{figure}
    \centering
    \begin{subfigure}[b]{0.45\textwidth}
        \includegraphics[width=\textwidth]{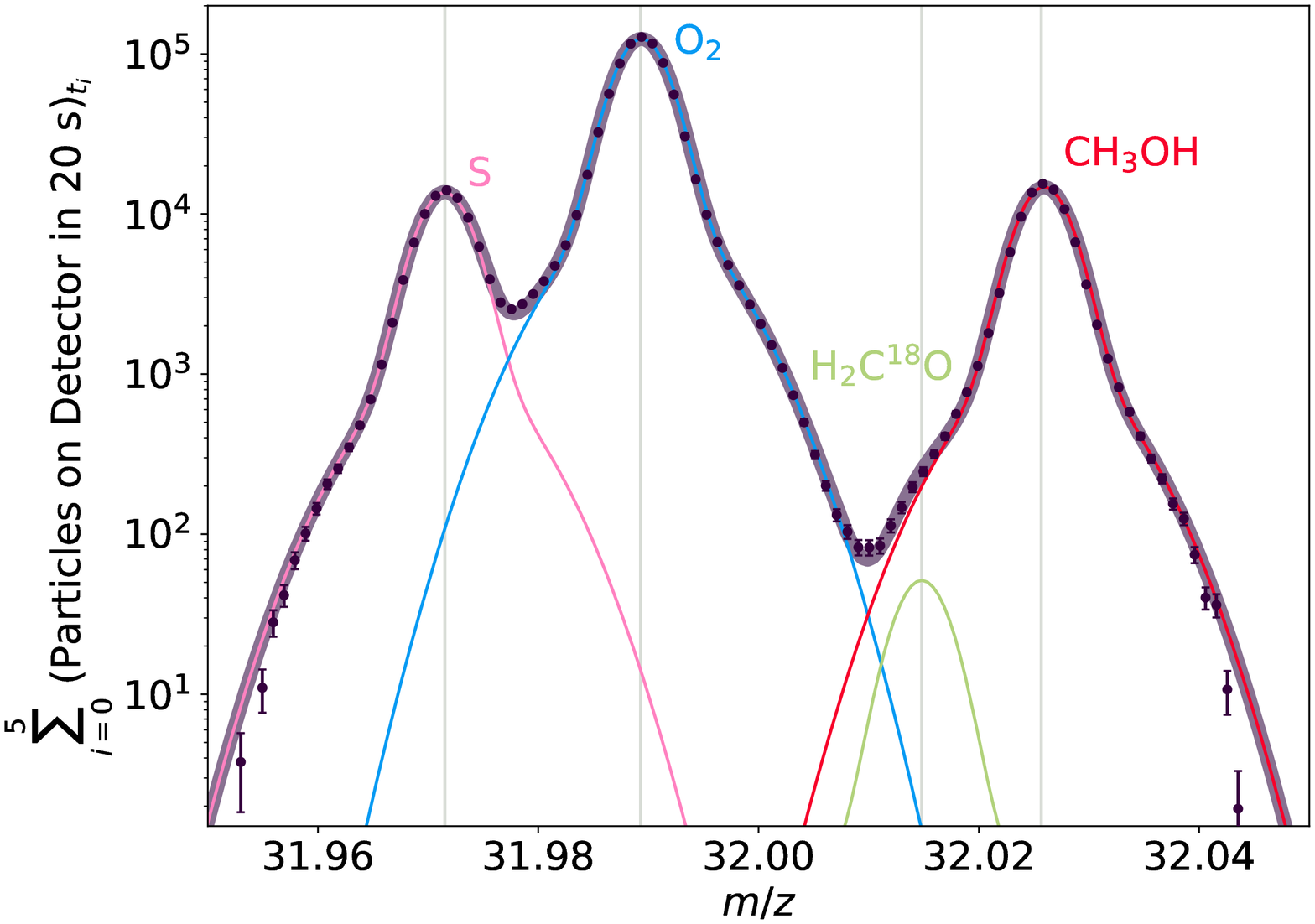}
        \caption{Normal methanol}
				\label{fig_ROSINA_32_O}
    \end{subfigure}
    ~ 
    \begin{subfigure}[b]{0.45\textwidth}
        \includegraphics[width=\textwidth]{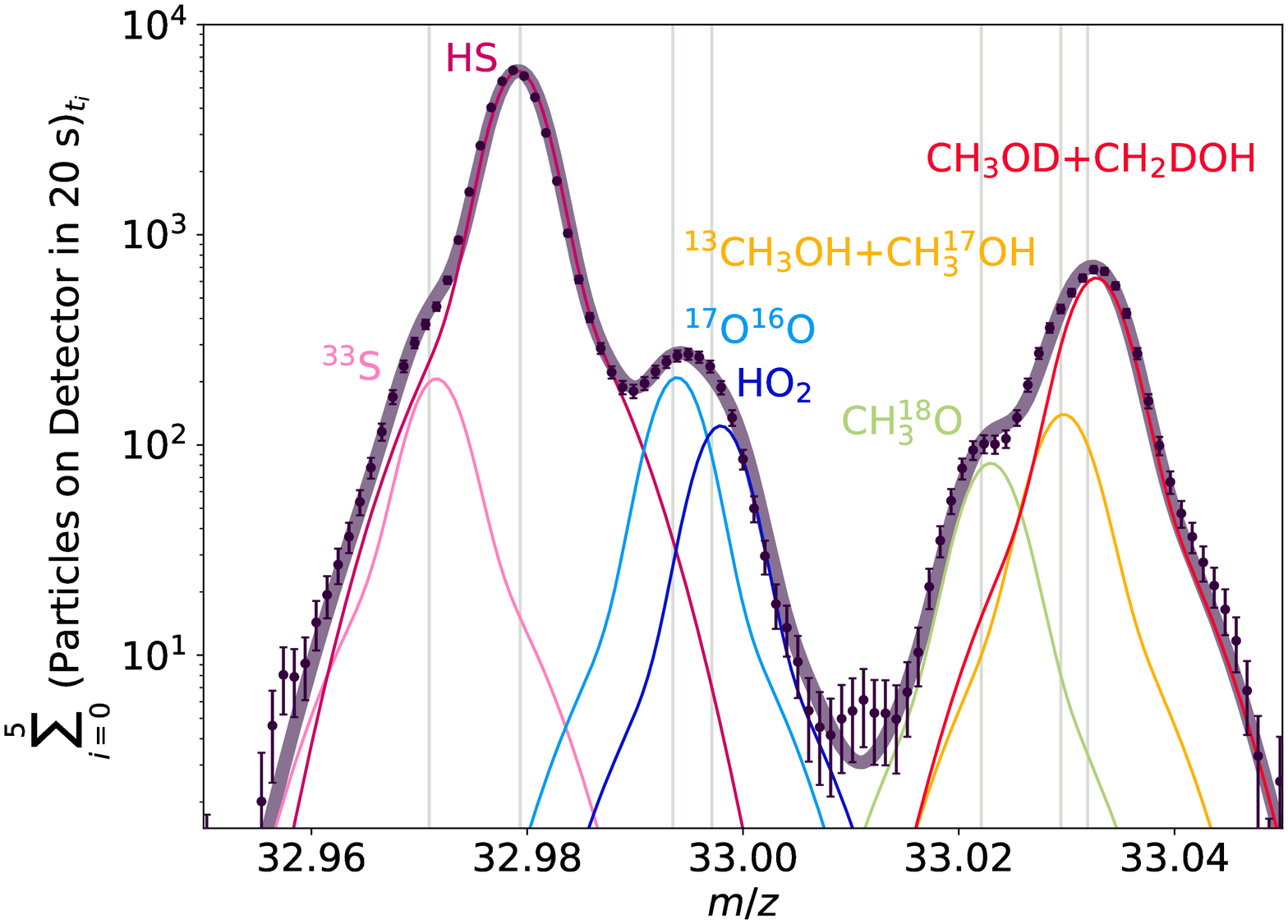}
        \caption{Mono-deuterated methanol}
				\label{fig_ROSINA_33_O}
    \end{subfigure}
		\begin{subfigure}[b]{0.45\textwidth}
        \includegraphics[width=\textwidth]{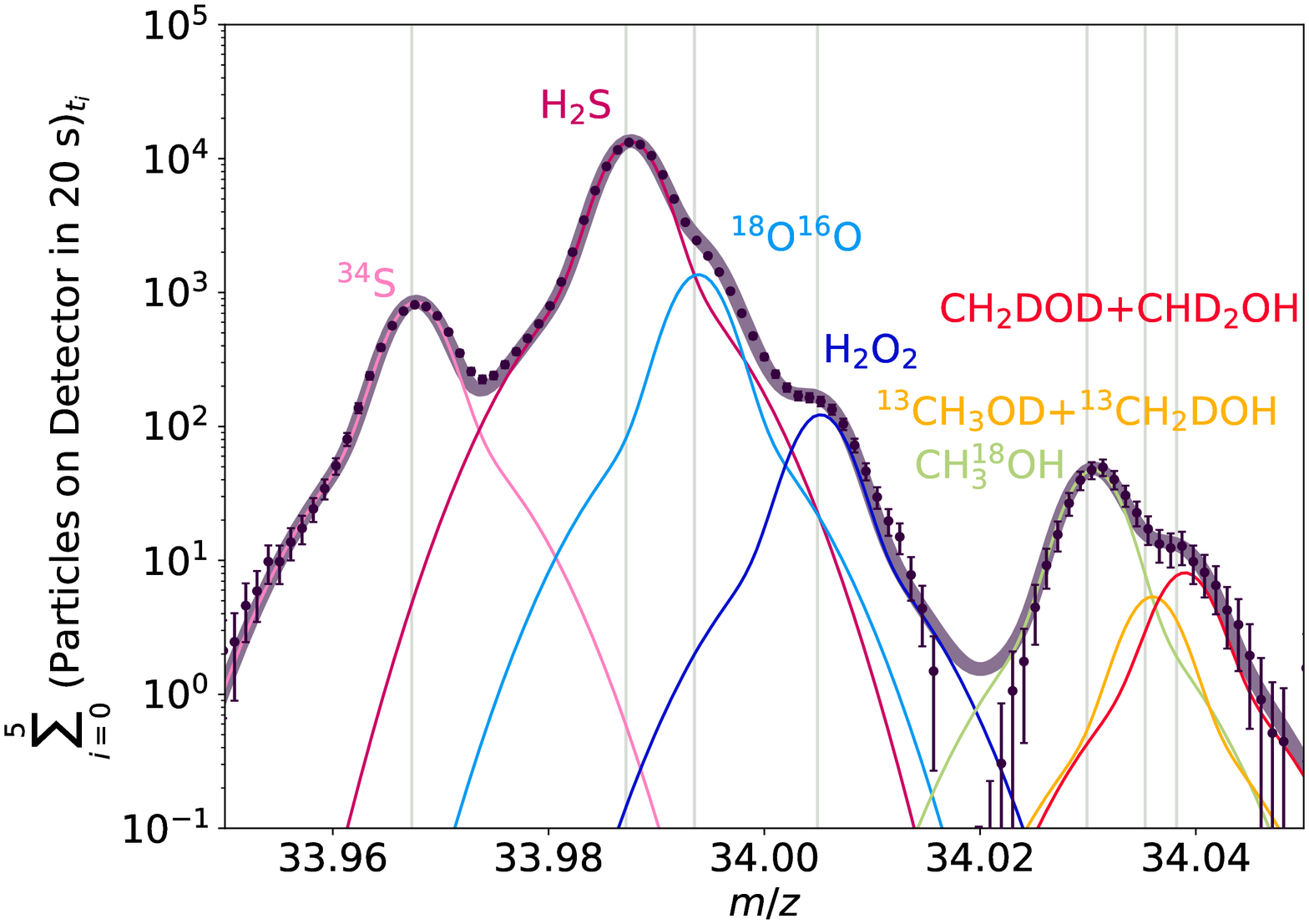}
        \caption{Di-deuterated methanol}
				\label{fig_ROSINA_34_O}
    \end{subfigure}
		\caption{ROSINA mass spectra for normal, mono-deuterated, and di-deuterated methanol on $m/z = 32, 33,$ and $34$, respectively, as measured on Row A and summed for the early October data set ($6$ in total: three packets on October 9th and three on the 19th, 2014) are shown as the dark data points. The depicted statistical error bars are smaller than the data points themselves when they are not visible. The individual contributors to the mass peaks are indicated with vertical lines at their exact masses, and associated double Gaussian fits are shown as thin colored curves. The thick dark purple curve is the sum of the individual double Gaussians and represents the overall fit to the measured ROSINA mass spectra.}
\end{figure}

This Appendix presents the ROSINA mass spectra for normal, mono-deuterated, and di-deuterated methanol at $m/z = 32$, $33$, and $34$, in Figs.~\ref{fig_ROSINA_32_O}--\ref{fig_ROSINA_34_O}, respectively, for the data set matching the spectra presented by \citet{Altwegg2020b} for the study of oxygen isotopologues, which is a sum of three packets on October 9th and three on the 19th, 2014 ($6$ in total; i.~e., the early October data set). There is only one minor difference between these spectra and those in \citet{Altwegg2020b}: the October 17th, 2014 spectrum is excluded from the analysis performed in this paper for deuterated methanol due to offset irregularities at the mass peak of di-deuterated methanol, which is not problematic for the study of $^{18}$O-methanol. The second data set of late October--December is presented in Section~\ref{mass_spec}. Table~\ref{tabl_ROSINA_OC} gives the measured methanol isotopologue abundance ratios and the oxygen and carbon isotopic ratios for the two data sets individually. Table~\ref{tabl_ROSINA_DH} tabulates the D/H ratio in methanol as calculated with the different assumptions based upon the average D-methanol/methanol and D$_{2}$-methanol/methanol ratios from the two studied data sets.

{\onecolumn
 \begin{center}
 \topcaption{Measured ROSINA methanol isotopologue abundance ratios relative to normal methanol for the early October and the late October--December 2014 data sets, and the corresponding oxygen and carbon isotopic ratios (with statistical error propagation).}
 \label{tabl_ROSINA_OC}
 \tablefirsthead{\hline \multicolumn{1}{l}{ } & \multicolumn{1}{l}{\textbf{early October}} & \multicolumn{1}{l}{\textbf{late October--December}} \\ \hline}
 \begin{xtabular*}{\textwidth}{l@{\extracolsep{\fill}}lll}
 CH$_{3}$OH                  & $1$                 & $1$\T\\
 $^{13}$CH$_{3}$OH           & $0.011\pm0.003$     & $0.011\pm0.003$\\
 CH$_{3}$OD$ + $CH$_{2}$DOH  & $0.047\pm0.011$     & $0.063\pm0.014$\\
 CH$_{3}^{18}$OH             & $0.0023\pm0.0008$   & $0.0021\pm0.0007$\\
 CH$_{2}$DOD$ + $CHD$_{2}$OH & $0.00060\pm0.00024$ & $0.00078\pm0.00032$\B\\
 \hline
 $^{16}$O/$^{18}$O           & $435\pm146$         & $476\pm160$\T\\
 $^{12}$C/$^{13}$C           & $91\pm22$           & $91\pm23$\B\\
 \hline
 \end{xtabular*}
 \end{center}
\twocolumn}

\section{Caveats of interstellar observations}
\label{caveats}

\begin{figure*}
 \centering
  \includegraphics[width=0.8\textwidth,height=0.8\textheight,keepaspectratio]{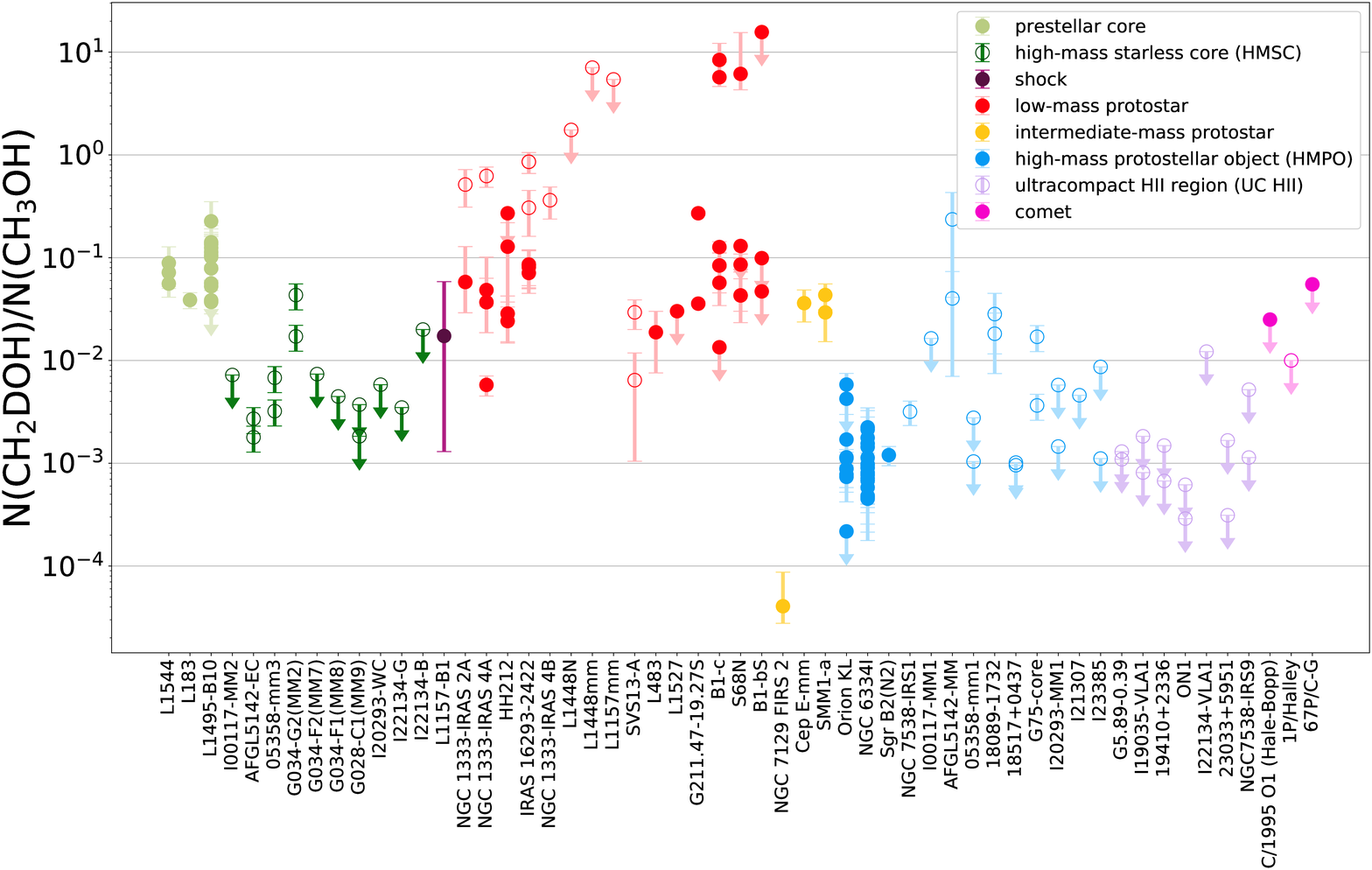}
 \caption{The ratio of CH$_{2}$DOH to CH$_{3}$OH towards star-forming regions and in comets, color-coded by the type of source (exhaustive version). Single dish observations of low-mass protostars and high-mass star-forming regions (not prestellar cores) have been indicated with open circles, while interferometric observations are marked by filled circles. The mass spectrometry measurement for comet 1P/Halley is also indicated with an open circle due to the complication by the required complex ion modeling (Section~\ref{cavCom}). Full references are given in Table~\ref{tabl_Dmeth}.}
 \label{fig_CH2DOH_all}
\end{figure*}

\begin{figure*}
 \centering
  \includegraphics[width=0.8\textwidth,height=0.8\textheight,keepaspectratio]{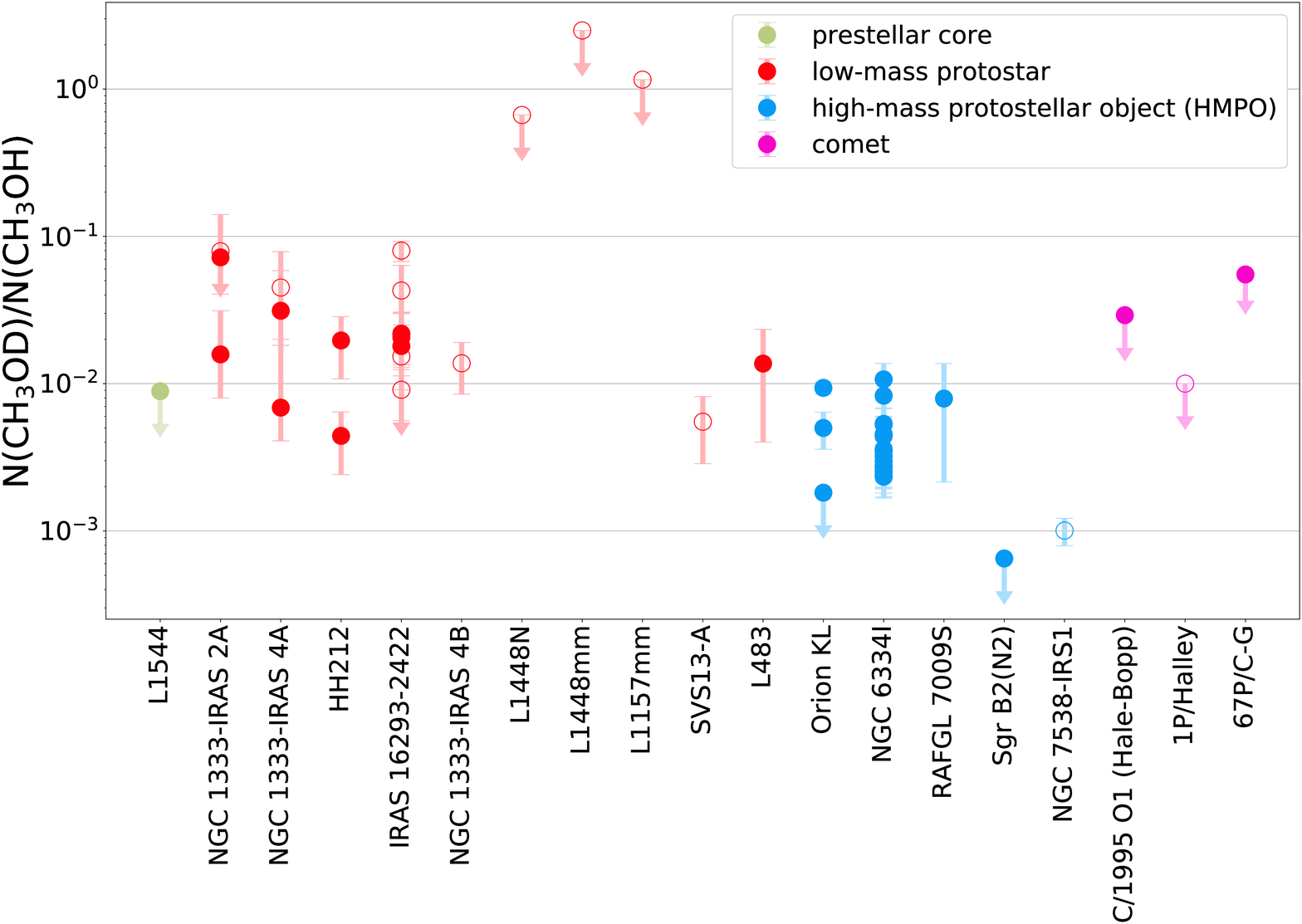}
 \caption{The ratio of CH$_{3}$OD to CH$_{3}$OH towards star-forming regions and in comets, color-coded by the type of source (exhaustive version). Single dish observations of low-mass protostars and high-mass star-forming regions (not prestellar cores) have been indicated with open circles, while interferometric observations are marked by filled circles. The mass spectrometry measurement for comet 1P/Halley is also indicated with an open circle due to the complication by the required complex ion modeling (Section~\ref{cavCom}). Full references are given in Table~\ref{tabl_Dmeth}.}
 \label{fig_CH3OD_all}
\end{figure*}

\begin{figure*}
 \centering
  \includegraphics[width=0.8\textwidth,height=0.8\textheight,keepaspectratio]{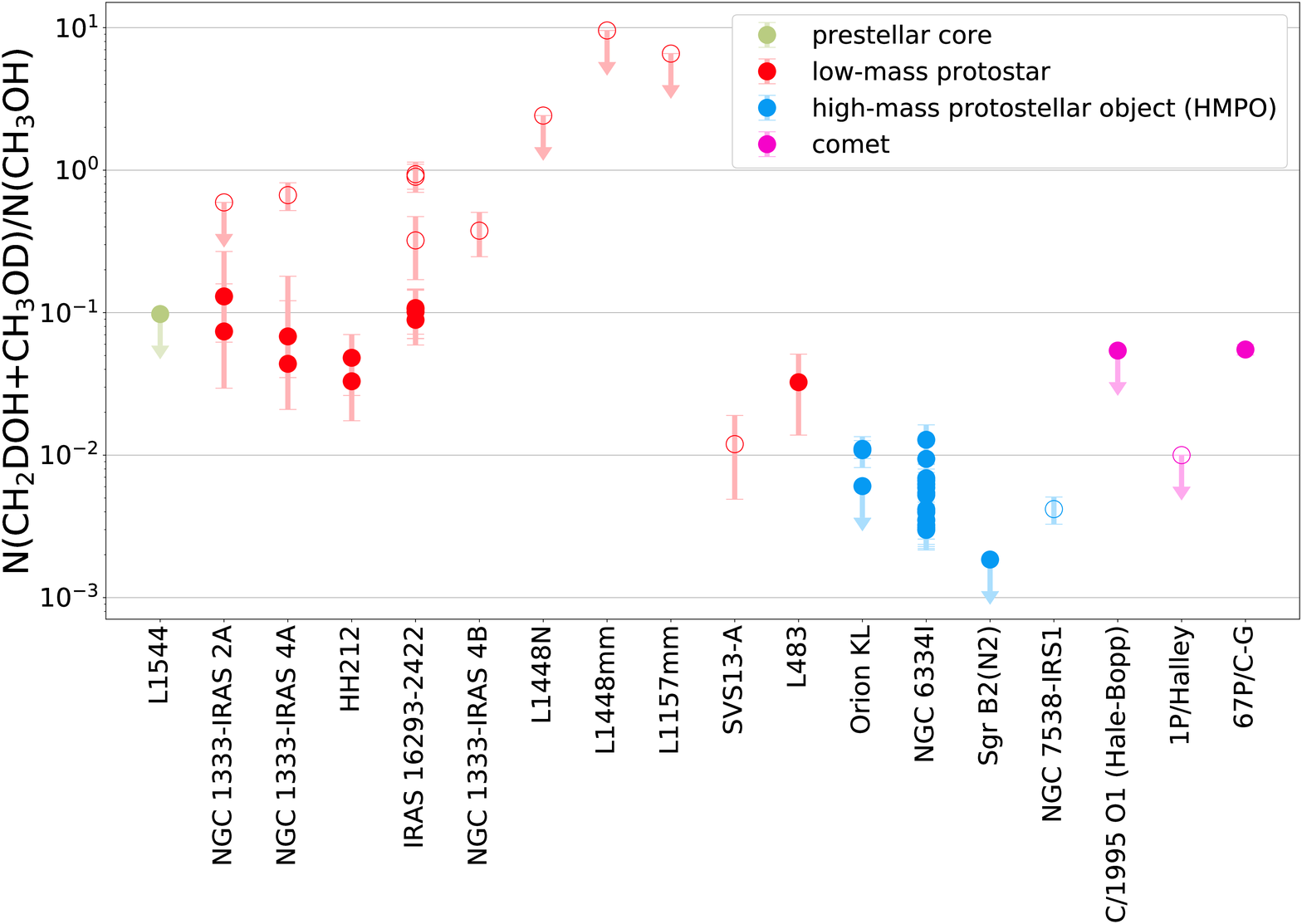}
 \caption{The ratio of CH$_{2}$DOH+CH$_{3}$OD to CH$_{3}$OH towards star-forming regions and in comets, color-coded by the type of source (exhaustive version). Single dish observations of low-mass protostars and high-mass star-forming regions (not prestellar cores) have been indicated with open circles, while interferometric observations are marked by filled circles. The mass spectrometry measurement for comet 1P/Halley is also indicated with an open circle due to the complication by the required complex ion modeling (Section~\ref{cavCom}). Full references are given in Table~\ref{tabl_Dmeth}.}
 \label{fig_Dmethtot_all}
\end{figure*}

Section~\ref{Dmeth} states that derivations of column densities of interstellar methanol and its isotopologues come with several caveats. These are discussed one-by-one in detail in the following subsections. Only the most-reliable interstellar methanol observations have been shown in Figs.~\ref{fig_CH2DOH}--\ref{fig_Dmethtot}, meanwhile an exhaustive comparison is shown in Figs.~\ref{fig_CH2DOH_all}--\ref{fig_Dmethtot_all}. References are tabulated in Table~\ref{tabl_Dmeth} with additional details provided in Appendix~\ref{quantities}. For low- and high-mass protostars, single dish observations have been indicated with open circles (specifically, these correspond to data reported in \citealt{vanDishoeck1995, Parise2002, Parise2004, Parise2006, Fontani2015, Bianchi2017a, Ospina-Zamudio2019a}). These observations are prone to beam dilution. The effects of this may cancel out if deuterated and non-deuterated methanol have the same spatial distribution, but this may not necessarily be the case. Moreover, depending on the specific publication, there are also large uncertainties stemming from disregarded dust and line optical depth effects and the use of poorly constrained partition functions for CH$_{3}$OD. From Figs.~\ref{fig_CH2DOH_all} and~\ref{fig_CH3OD_all} it can be seen that for sources with follow-up interferometric observations (solid points), the ratio of CH$_{2}$DOH to CH$_{3}$OH decreases by roughly one order of magnitude and the ratio of CH$_{3}$OD to CH$_{3}$OH decreases by a factor of a few in almost all instances in comparison to the values derived based on single dish data. Figs.~\ref{fig_CH2DOH_all}--\ref{fig_Dmethtot_all} do allow the comparison to be extended to the birth places of high-mass protostars, i.~e., high-mass starless cores (HMSCs). HMSCs tend to have a CH$_{2}$DOH/CH$_{3}$OH ratio that is an order of magnitude lower than prestellar cores, i.~e., in the range of  $\sim10^{-2} - 10^{-3}$. The later stage of ultracompact HII regions (UC HIIs) spans a range of CH$_{2}$DOH/CH$_{3}$OH ratios that agrees with that covered by HMPOs.

\subsection{Spatial distribution}

A fundamental caveat of observations of distant objects such as prestellar cores and protostars is the uncertainty in the spatial extent of the emission, which can vary from molecule to molecule and even from line to line towards a single target. In prestellar cores, methanol and its isotopologues are most abundant in the solid phase due to the low dust temperature of these environments. The gas-phase observations of methanol in prestellar cores discussed in this work are primarily a result of non-thermal desorption processes caused by cosmic rays (Section~\ref{meth_chem}), because of methanol's high binding energy ($E_{\text{des}} \approx 4930$~K; \citealt{BrownBolina2007}). Differences in spatial distribution of methanol in comparison to that of its isotopologues may be explained either through variations in the underlying solid phase abundances or differing efficiencies of non-thermal desorption depending on the methanol flavor. For example, if D-methanol formation is occurring more frequently than non-deuterated methanol formation, then reactive desorption may favor the release of deuterated rather than non-deuterated methanol into the gas (assuming the same reactive desorption efficiency for both formation reactions). Gas-phase chemistry is unlikely to alter the spatial extent of methanol and its isotopologues in prestellar cores due to the low densities. Single dish observations may not adequately pick up clumpy spatial distribution of methanol and its isotopologues in a core, which has been claimed for L1544 and L183 based on IRAM~30m observations \citep{Bizzocchi2014, Chacon-Tanarro2019, Lattanzi2020}. However, interferometric observations struggle to recover the most extended structures of cores \citep{Caselli2019}.

Protostars are morphologically much more complex than prestellar cores due to the presence of collapsing envelopes, rotating discs, and bipolar outflows on spatial scales that vary from tens of au to thousands of au. Gaseous methanol emission from protostars is dominated by thermally desorbed methanol in hot regions of high density, although non-thermally desorbed methanol in cold zones is still present in the low-density enshrouding envelope. Methanol and its deuterated isotopologues are expected to thermally desorb in the same regions as the mass difference between the various flavors is small \footnote{The thermal desorption rate is proportional to $m^{-0.5}$, where $m$ is the molecular mass. Mono-deuterated methanol is heavier by $3.1$ per cent than CH$_{3}$OH, di-deuterated by $6.3$ per cent, tri-deuterated by $9.4$ per cent, and fully deuterated by $13$ per cent.}. Consequently, variations in the spatial distribution must stem either from an underlying difference in solid methanol or from active on-the-spot gas-phase chemistry (which is possible due to the higher temperatures and densities). Single dish observations can only disentangle different components based on velocity for data sets with high spectral resolution. For example, for the case of L483, the IRAM~30m survey of \citet{Agundez2019} showed that most of the detected CH$_{3}$OH emission comes from the ambient core with only a minor contribution from the outflow based on the velocity profiles of the lines. Meanwhile, the narrower lines of deuterated molecules suggest their presence solely in less turbulent, colder regions of the envelope. Alternatively, if a sufficiently large frequency range is targeted and a larger number of lines are detected, components can be distinguished based on the derived excitation temperatures, as for the low-mass Class I source SVS13-A \citep{Bianchi2017a}. Interferometric observations have the possibility of spatially disentangling the various components of such a system, or at least to separate individual components of a binary system. With interferometric Atacama Large Millimeter/submillimeter Array (ALMA) observations, IRAS~16293-2422~A could be studied separately from~B to derive similar (at most a factor of $1.2$ apart) CH$_{2}$DOH/CH$_{3}$OH ratios towards both protostars \citep{Jorgensen2018, Manigand2020a}. For NGC~1333-IRAS~4A, interferometric IRAM Plateau de Bure Interferometer (IRAM-PdBI; now called the NOrthern Extended Millimeter Array, NOEMA) observations determined a CH$_{2}$DOH/CH$_{3}$OH ratio of $0.037$ \citep{Taquet2019}. Subsequent interferometric ALMA observations separated the source into its binary components A1 and A2 with ratios of $0.0058$ and $0.048$, respectively \citep{Sahu2019}, indicating that the two binary components are starkly different from one another (and that earlier IRAM-PdBI observations were dominated by the emission from A2). In protostellar regions, methanol is likely present in the extended cold circumstellar/circumbinary envelope, as well as on the most inner hot regions in the vicinity of the protostar(s). Consequently, care must be taken when interpreting interferometric observations, which may suffer from spatial filtering of extended structures already upon the increase of spatial resolution from arcseconds to subarcseconds.

Interpretation of observations of intermediate- and high-mass star-forming regions is even trickier due to the larger distances involved and the consequent difficulty in spatially resolving merely the number of protostars in a source. In the binary intermediate-mass system Cep~E-mm, it seems that there are three physical components traced by methanol depending on the upper energy level and the column density obtained from the rarer isotopologue differs by an order of magnitude from that derived based on optically thin lines of methanol \citep{Ospina-Zamudio2018}. Clumpy emission from methanol and its deuterated isotopologues is also reported for the massive Orion Becklin-Neugebauer/Kleinmann-Low (BN/KL) star-forming region \citep{Peng2012}. Future interferometric observations at even higher spatial resolution may show additional spatial segregation of methanol and its isotopologues.

\subsection{Excitation}

Most column densities of molecules in prestellar cores and towards protostars are derived based upon the local thermal equilibrium (LTE) assumption. Particularly under the low densities of cores, this assumption may no longer be valid. However, analysis of data on L1544 and L183 are showing differences of only a factor of $2$ between LTE and non-LTE CH$_{3}$OH column densities \citep{Bizzocchi2014, Lattanzi2020}. Non-LTE calculations still suffer from a lack of collisional data for many species, because either too few collisional partners are available in databases \citep{Bizzocchi2014}, or not all excited levels are taken into account \citep{Neill2013b}. For CH$_{2}$DOH, no collisional coefficients are available at all, making it impossible to perform non-LTE calculations \citep{Parise2006, Codella2012}. The same situation holds for CH$_{3}$OD. It has been speculated that non-LTE effects may be non-negligible for this isotopologue due to high critical densities for some of its transitions (based on observations of low-mass protostars NGC~1333-IRAS~2A and~-IRAS~4A; \citealt{Taquet2019}). Low-level non-LTE effects were also thought to explain partially the scatter in the methanol deuterated isotopologue rotational diagrams from observations of IRAS~16293-2422 \citep{Parise2002, Parise2006}; however, such scatter is more likely explained by dust optical thickness and both components of the binary being probed jointly in the earlier IRAM~30m observations \citep{Jorgensen2016}. For $^{13}$CH$_{3}$OH, LTE and non-LTE calculations have also been demonstrated to agree closely for the low-mass protostars SVS13-A and HH212, i.~e., regions that are denser than the earlier prestellar phases \citep{Bianchi2017a, Bianchi2017b}. Under the LTE assumption, for protostars, it appears that the observed emission spectra are most sensitive to the column density rather than the excitation temperature. In the past, it was difficult to constrain the excitation temperature due to a limited frequency coverage of observations and consequently, a narrow range of upper energy levels being probed in a single data set. However, the poorly constrained excitation temperature is not expected to affect ratios of column densities of isotopologues of the same molecule significantly. Nowadays, with the wider frequency coverage of ALMA, many more lines are probed towards protostars, allowing an accurate determination of the excitation temperature. The situation remains trickier for starless and prestellar cores, as only low upper energy levels of a molecule are typically excited, leaving few lines for the derivation of the excitation temperature. Overall, uncertainties stemming from the LTE assumption are likely minor for all regions of star-forming systems.

\subsection{Optical thickness}

Another critical parameter is that of optical depth, which is set by the gas and the dust along the line of sight. In prestellar cores, the densities of gas and dust are low enough for the emission to be typically optically thin even for the most abundant methanol CH$_{3}$OH variant \citep{Bizzocchi2014, Chacon-Tanarro2019, Lattanzi2020}. In protostars, densities are much higher and thus, optical depth becomes a much graver concern. CH$_{3}$OH gas is thought to always be optically thick (e.~g., \citealt{Menten1988, Parise2006, Neill2013b, Fuente2014, Jorgensen2016, Bianchi2017a, Bianchi2017b, Jorgensen2018, Bogelund2018, Ospina-Zamudio2018, Taquet2019, LeeChin-Fei2019a, Sahu2019, Agundez2019, Manigand2020a}). For some sources, even the singly deuterated isotopologues (as for on-source in NGC~1333-IRAS~2A and -IRAS~4A for CH$_{2}$DOH and CH$_{3}$OD, \citealt{Taquet2019}; or as for CH$_{2}$DOH already at a half-$0.5\arcsec$-beam offset position from source B in IRAS~16293-2422, \citealt{Jorgensen2018}) and the $^{13}$C-isotopologue may be (partially) optically thick (as in NGC~6334I, \citealt{Bogelund2018}; or in Sagittarius~B2(N2), \citealt{Muller2016}; or at a half-$0.5\arcsec$-beam offset position from source B in IRAS~16293-2422, \citealt{Jorgensen2016, Jorgensen2018}). The $^{18}$O isotopologue is thought to be optically thin and, hence, be a reliable indicator of the abundance of methanol. However, CH$_{3}^{18}$OH observations require either a source that is bright in methanol emission or data of sufficiently high sensitivity. Unfortunately, methanol column density towards protostars is approximated based upon this most-reliable isotopologue in very few cases (namely: NGC~6334I, \citealt{Bogelund2018}; IRAS~16293-2422~A and~B, \citealt{Jorgensen2016, Jorgensen2018, Manigand2020a}; B1-c and S68N, \citealt{vanGelder2020b}; SMM1-a, Ligterink et al. 2020d in prep.). As the gas-to-dust ratios may vary across star-forming regions, and also between individual components of such systems, it is not always obvious when dust opacity starts to play a role in a frequency-dependent fashion ($\tau_{\nu} \propto \nu^{\beta}$, where $\nu$ is the frequency and $\beta$ is the dust opacity exponent). Severe examples of dust optical thickness at submm ALMA Band~7 frequencies to the point that the majority of observed lines appear in absorption were seen on-source B in IRAS~16293-2422 \citep{Jorgensen2016} and on-source in NGC~1333-IRAS~4A1 \citep{Sahu2019}. Dust extinction has also been reported to play a role in Orion~KL \textit{Herschel}/HIFI observations \citep{Neill2013b}. Observations suffering from gas and/or dust optical thickness effects may be underestimating the column density of methanol by an order of magnitude, and by factors of a few of its isotopologues. Optical depth is likely the cause of the largest uncertainties in methanol observations of star-forming regions.

When an optically thin isotopologue is used to derive the column density of normal methanol, the adopted isotopic ratio starts to be of importance as well. In Figs.~\ref{fig_CH2DOH}--\ref{fig_Dmethtot}, for the case of the $^{13}$C-isotopologue being used to derive the methanol column density, the adopted $^{12}$C/$^{13}$C ratios span a range from $50$ \citep{LeeChin-Fei2019a} to $77$ \citep{Codella2012, Fontani2015}. For the case of the $^{18}$O-isotopologue, the $^{16}$O/$^{18}$O ratio spans from $450$ \citep{Bogelund2018} to $560$ \citep{Jorgensen2018}. This corresponds directly to differences by factors of $1.54$ and $1.24$, respectively, in the methanol column densities depending on the adopted isotopic ratio. These variations are due to the galactocentric distance of the targeted sources; however, the relations prescribing the isotopic ratios with galactocentric distance are hard to determine accurately and may vary from molecule to molecule (e.~g., \citealt{Milam2005, Jacob2020}). Table~\ref{tabl_COisoratios} contains a complete list of the isotopic ratios employed in respective publications that were used to obtain Figs.~\ref{fig_CH2DOH}--\ref{fig_Dmethtot}.


\subsection{Spectroscopic uncertainties}

Methanol and its isotopologues are asymmetric top rotors with complex torsion-vibrational coupling \citep{SibertCastillo-Chara2005}. For the normal isotopologue and the $^{18}$O isotopologue, the partition functions have been determined in a dedicated molecule-tailored fashion by summing rovibrational states (for $v_{\text{t}}\leq3$; $J\leq44$; $K\leq20$ for both) with the spectroscopic data covering $v_{\text{t}}\leq2$, $J_{\text{max}}=40$, $K_{\text{max}}=20$ and $v_{\text{t}}\leq2$, $J_{\text{max}}=30$, $K_{\text{max}}=15$, respectively. For the $^{13}$C isotopologue, the spectroscopic data cover $v_{\text{t}}\leq1$, $J_{\text{max}}=20$, $K_{\text{max}}=14$ with the partition function being derived by scaling that of the $^{18}$O isotopologue (which deviated only slightly from that derived by scaling the partition function of normal methanol). These three species are available in the Cologne Database of Molecular Spectroscopy (CDMS; \citealt{Muller2001, Muller2005, Endres2016}) with further details in the online documentation and the references therein. For CH$_{2}$DOH, a spectroscopic entry exists in the Jet Propulsion Laboratory (JPL) catalogue \citep{Pickett1998} covering the three ground sub-states $e_{0}$, $e_{1}$, and $o_{1}$ for $J_{\text{max}}=40$, $K_{\text{max}}\sim9$ and the partition function being calculated by summing over the rotational states as given in \citet{Pearson2012}. For CH$_{3}$OD, no publicly available entry exists in neither the CDMS nor the JPL databases. The thus-far published observational works have employed self-compiled line lists and differing assumptions on the partition function. The preferred methodology is that of \citet{Belloche2016, Jorgensen2018, Bogelund2018}, which approximates the partition function of CH$_{3}$OD by scaling that of CH$_{3}^{18}$OH. This approach is expected to yield line strength uncertainties of $\sim5-10$ per cent (based on an analogous exercise being carried out for CH$_{3}$SD, which resulted in uncertainties of only $1$ per cent; priv. comm. H.~S.~P.~M{\"u}ller). Earlier works have also derived the partition function of CH$_{3}$OD based on the rigid rotor approximation. \citet{Taquet2019} showed that the use of these two different assumed partition functions can lead to a factor of $5$ difference in the column density of CH$_{3}$OD. Methanol is far from a rigid rotor, therefore this approximation should be avoided (if possible) for CH$_{3}$OD and CH$_{2}$DOH. Dedicated spectroscopic characterization of deuterated methanol isotopologues is urgently needed.


When a molecule is being considered as a rigid object, it is assumed that the bond lengths are fixed and that the molecule cannot vibrate. In reality, a molecule vibrates as it rotates and the bonds are elastic rather than rigid. However, the rigid-rotor approximation is fair, because the amplitude of the vibration is small compared to the bond length. The effect of centrifugal stretching is smallest at low $J$ values, and so its effect is smallest at low temperatures. However, at higher temperatures, vibrational correction factors are applied by some authors. For CH$_{2}$DOH, \citet{Taquet2019} applied a vibrational correction factor of $1.15$ at $160$~K (also used by \citealt{Belloche2016}) and of $1.46$ at $300$~K (also used by \citealt{Jorgensen2018}) based on the torsional data of \citet{Lauvergnat2009}. For CH$_{3}$OD, \citet{Bogelund2018} applied a vibrational correction factor of $1.25$ for a $\sim120-340$~K temperature range and \citet{Belloche2016} used $1.05$ at $160$~K. Column densities are divided by these vibrational correction factors directly. Since the spectroscopic entries for CH$_{3}$OH, $^{13}$CH$_{3}$OH, and CH$_{3}^{18}$OH account for vibrational contribution up to $v_{\text{t}}=1-2$, further vibrational correction factors to the column densities are not needed at $160$~K \citep{Muller2016}. Although there are limitations in terms of the $K_{\text{max}}$ and, more severely, $J_{\text{max}}$ considered, uncertainties in the line strengths should be minor at $\sim200$~K and in the $10-20$ per cent range at $\sim300$~K (priv. comm. H.~S.~P.~M{\"u}ller).


\section{Deuteration of methanol}
\label{meth_Dchem}

\subsubsection{Deuteration in the methyl functional group}

In star-forming regions, D atoms are available alongside H atoms for grain-surface chemistry. Methanol deuteration reaction schemes depend on the order in which deuteration is to occur. If CH$_{3}$OH is formed before deuterated methanol, then D-methanol can be made via H-D substitution reactions. This was experimentally determined by bombarding CH$_{3}$OH ice with D atoms at $10$~K, which also showed that H-D substitution reactions produce only CH$_{2}$DOH and not CH$_{3}$OD in this scenario \citep{Nagaoka2005}. Once deuterated, methanol is thought to keep its deuterium with hydrogenation being inhibited \citep{Nagaoka2005}. D-methanol could potentially be deuterated further via H-D substitution reactions if exposed to D atoms, but this deuteration is highly inefficient \citep{Hiraoka2005}.

If deuteration was to occur following formaldehyde formation, then only D$_{4}$-methanol would be produced. Experiments of \citet{Hidaka2009} show that exposure of H$_{2}$CO ice to D atoms does not produce D- nor D$_{2}$-methanol. HDCO, D$_{2}$CO, and CD$_{3}$OD were the sole products of this via H-D substitution reactions and subsequent D atom additions to D$_{2}$CO to form fully deuterated methanol. This implies that H-D substitution reactions are faster than D additions, but may occur if H-D substitution is no longer possible (i.~e., when formaldehyde is fully deuterated in the absence of competing H atom additions). However, if formaldehyde is formed, then deuterated via H-D substitution reactions, and then hydrogenated further to methanol, CHD$_{2}$OH can form. Experiments exposing D$_{2}$CO ice to H atoms result in CHD$_{2}$OH via H atom additions \citep{Hidaka2009}. These experiments also show that H and D addition to HDCO is not competitive with other simultaneous processes, and D-methanol is not produced.

If deuteration was to occur simultaneously with hydrogenation starting from CO, all varieties of deuterated methanol can be produced except for CH$_{3}$OD and CH$_{3-n}$D$_{n}$OD for $n=1-3$ as shown in the laboratory \citep{Nagaoka2005}. The suite of executed laboratory work implies that CD$_{3}$OD would be evidence of an extreme prevalence of D over H atoms after the formation and deuteration of formaldehyde. CD$_{3}$OH and CHD$_{2}$OH are products of simultaneous hydrogenation and deuteration starting from CO and after formaldehyde formation. CH$_{2}$DOH points to continuation of D atom availability after the formation of methanol (without contributing channels via formaldehyde and inefficient conversion to more deuterated methanol isotopologues). Detections of CH$_{2}$DOH, CHD$_{2}$OH, and CD$_{3}$OH towards the low-mass protostar IRAS~16293-2422 provide observational support that D atoms remain available on grain surfaces after methanol formation; and that hydrogenation and deuteration initiate hand in hand starting from CO. The up-to-now non-detected CD$_{3}$OD is likely an indicator of the expected prevalence of H atoms over D atoms after formaldehyde formation.


\subsubsection{Deuteration in the hydroxyl functional group}

CH$_{3}$OD has not been formed in any of the laboratory experiments involving CO, H$_{2}$CO, CH$_{3}$OH, H, and D atoms. It is possible that it is below the detection limit of these experiments \citep{Nagaoka2007}; however, this would still imply an extremely low abundance of this isotopologue. When H reacts with CH$_{3}$OH (an abstraction reaction yielding H$_{2}$ and a radical), it may theoretically lead to either the methoxy (CH$_{3}$O) or the hydroxymethyl (CH$_{2}$OH) radical. Theoretical studies indicate that the methoxy channel is endothermic, while the hydroxymethyl channel is exothermic and has a lower vibrationally adiabatic barrier than that of methoxy \citep{KerkeniClary2004, Nagaoka2007}. Furthermore, the dissociation energy of CH$_{3}$OH to CH$_{3}$O+H is larger than that of CH$_{3}$OH to CH$_{2}$OH+H \citep{Bauschlicher1992}. Consequently, H abstraction proceeds faster from the methyl group versus that from the hydroxyl group, and H-D substitution reactions (H abstractions followed by D atom additions) produce CH$_{2}$DOH almost exclusively. The dominance of the CH$_{2}$DOH channel over that of CH$_{3}$O is also supported by the experiments analyzing simultaneous complex organic molecule formation upon radical-radical associations \citep{Chuang2016}. These findings have also been verified with quantum chemical computations \citep{GoumansKaestner2011}. H-D substitution may proceed through an excited intermediate species CH$_{3}$DOH$^{*}$; however, this is unlikely to have a significant contribution due to a higher theoretically calculated direct H-H exchange reaction rate barrier \citep{Osamura2004, Nagaoka2007}. Formation pathways of CH$_{3}$OD starting from deuterated isotopologues of formaldehyde are excluded as such products will retain at least one deuterium atom in the methyl group, as H-D substitution reactions are faster than D additions. CH$_{3}$OD may form starting from non-deuterated formaldehyde and the CH$_{3}$O radical (that is preferentially produced by H$_{2}$CO+H), but likely at a very low rate as H addition proceeds more efficiently than that of D.

In star-forming regions, CH$_{3}$OD is nevertheless a frequently detected molecule (Section~\ref{Dmeth}), which needs an explanation. It is possible that the long astronomical time-scales may produce detectable abundances via all the minor formation channels combined. Potentially, in conjunction with effects of cosmic rays and protostellar UV irradiation (upon the emergence of a protostar), the availability of the CH$_{3}$O radical may be higher for a longer amount of time, allowing additional formation of CH$_{3}$OD by D atom addition. The dominant methanol photodissociation channel is still under debate; however, CH$_{3}$O is not expected to be the dominant fragment \citep{Hagege1968, Oberg2009, Laas2011, Bertin2016}. Consequently, CH$_{3}$O is expected to only be available in the synthesis chain towards CH$_{3}$OH. The internal rearrangement of CH$_{2}$DOH into CH$_{3}$OD has been ruled out due to very high barriers for such an exchange reaction for neutral, ionic, and protonated forms of methanol \citep{Osamura2004}. Alternatively, CH$_{3}$OD may form upon an isotope exchange reaction between non-deuterated methanol and deuterated water or deuterated ammonia. Laboratory experiments have shown that the hydroxyl group of deuterated methanol undergoes H-D exchange reactions with non-deuterated water due to its ability to hydrogen bond unlike the methyl group \citep{Souda2003, Souda2004, Ratajczak2009, FaureM2015}. Further experiments have demonstrated H-D exchange reactions taking place between deuterated and non-deuterated water molecules \citep{DevlinBuch2007, Galvez2011, Lamberts2015}. Such reactions have been suggested to alter the abundance of CH$_{3}$OD and to equilibrate its D/H ratio with that of water \citep{Ratajczak2011, FaureA2015, CeccarelliPPVI}; however, they could also be responsible for forming CH$_{3}$OD from mono- or di-deuterated water in the first place. Such a dedicated laboratory experiment has not been carried out yet, but must be undertaken and explored for a variety of temperatures, as such H-D exchange reactions have been shown to be sensitive to the phase of the water ice \citep{Lamberts2015}. This idea is supported by the time-of-flight secondary-ion mass spectrometry (TOF-SIMS) experiments of \citet{Kawanowa2004} with CH$_{3}$OH and D$_{2}$O ices, in which a clear peak is detected at the mass of D$^{+}$(CH$_{3}$DO). Although the applied technique cannot conclusively distinguish CH$_{3}$OD from CH$_{2}$DOH, fig.~$1$ of \citet{Kawanowa2004} does not show a peak at $m/z = 36$, which would indicate the presence of D$_{2}$-methanol. This experimentally suggests that H-D exchange reactions are only happening between the hydroxyl groups of water and methanol.

\section{Quantities of methanol and its isotopologues from star-forming regions to comets}
\label{quantities}

Quantities of methanol and its isotopologues from star-forming regions to comets used to construct Figs.~\ref{fig_CH2DOH}--\ref{fig_Dmethtot} and~\ref{fig_CH2DOH_all}--\ref{fig_Dmethtot_all} are tabulated in Table~\ref{tabl_Dmeth} alongside their respective references. The full descriptions of the probed locations in each source can be found in the corresponding publications, as well as the beam sizes of the observations, and assumptions about the source sizes. Methanol and its isotopologues are always considered to be co-spatial, therefore these assumptions cancel out when taking the ratio of methanol isotopologues relative to normal methanol. When either $^{13}$CH$_{3}$OH or CH$_{3}^{18}$OH is specified in brackets in the ``Position'' column of the table, that is an indicator that an isotopologue was used to determine the column density of normal methanol that is given in the table. The isotopic ratios used in the various publications are given in Table~\ref{tabl_COisoratios}. The largest differences stem from the variation of these ratios with galactocentric distance (e.~g., as in the case of NGC~6334I, \citealt{Bogelund2018}). The error bars in Figs.~\ref{fig_CH2DOH}--\ref{fig_Dmethtot} and~\ref{fig_CH2DOH_all}--\ref{fig_Dmethtot_all} match those reported in the respective works. In the case of NGC~6334I, the errors on the column densities are thought to be around $20$ per cent (priv. comm. N.~F.~W.~Ligterink). However, the given range in the paper is always larger, except for CH$_{3}^{18}$OH in MM1~II, for which an error of $20$ per cent is adopted. Such an error is also adopted if a range is not explicitly stated in \citet{Bogelund2018}. For L1157-B1, the published error for CH$_{2}$DOH of \citet{Codella2012} is taken in the positive direction, and the error in the negative direction is assumed to correspond to a column density that is an order of magnitude lower (priv. comm. C.~Codella). For L1157mm, the errors quoted in \citet{Parise2006} are unphysical, and thus the data point is indicated as an upper limit. For all the data points stemming from \citet{Fontani2015}, an error of $20$ per cent is assumed on all column densities, which is thought to be a reasonable estimate (priv. comm. F.~Fontani). For SVS13-A, an error of $20$ per cent was assumed, as no errors were described in the publication of \citet{Bianchi2017a}. When a reference is made to a temperature in K in the ``location'' column of the table, it refers to an assumption that has been made about the adopted excitation temperatures, see respective publications for details. P04 and J18 refer to two different partition functions for CH$_{3}$OD used in \citet{Taquet2019}, see Appendix~\ref{caveats}.

The upper limit of D-methanol/methanol in comet 1P/Halley as measured by \textit{Giotto}--NMS stems from \citet{Eberhardt1994}. The \textit{Rosetta}--ROSINA measurement of this value in comet 67P/C--G is described in Section~\ref{results}. Both of these measurements stem from mass spectrometry and were consequently not listed in Table~\ref{tabl_Dmeth}, but do appear in Figs.~\ref{fig_CH2DOH}--\ref{fig_Dmethtot} and~\ref{fig_CH2DOH_all}--\ref{fig_Dmethtot_all}.

{\onecolumn
 \begin{center}
 \topcaption{Carbon and oxygen isotopic ratios used in the literature for derivation of methanol column densities and used to construct Figs.~\ref{fig_CH2DOH}--\ref{fig_Dmethtot} and~\ref{fig_CH2DOH_all}--\ref{fig_Dmethtot_all}, including variations based on the galactocentric distance of the sources. The sources and the distances adopted for them in the respective publications are also provided when stated therein.}
 \label{tabl_COisoratios}
 \tablefirsthead{\hline \multicolumn{1}{l}{ } & \multicolumn{1}{l}{\textbf{Isotopic ratio}} & \multicolumn{1}{l}{\textbf{Source}} & \multicolumn{1}{l}{\textbf{Distance (pc)}} & \multicolumn{1}{l}{\textbf{Reference}} \\ \hline}
 \begin{xtabular*}{\textwidth}{l@{\extracolsep{\fill}}llll}
                  $^{12}$C/$^{13}$C~~ & $50$  & HH212            & $400$         & \citet{LeeChin-Fei2019a}\T\\
                                      & $60$  & Orion KL         & -             & \citet{Neill2013b}\\
                                      & $62$  & NGC 6334I        & $1~300$       & \citet{Bogelund2018}\\
                                      & $65$  & NGC 7129 FIRS 2  & $1~250$       & \citet{Fuente2014}\\
                                      & $68$  & SVS13-A          & $235$         & \citet{Bianchi2017a}\\
                                      &       & L483             & $200$         & \citet{Agundez2019}\\
                                      &       & IRAS 16293-2422  & $140$         & \citet{Manigand2020a}\\
                                      &       & SMM1-a           & $436.0\pm9.2$ & Ligterink et al. 2020d in prep.\\
                                      & $70$  & IRAS 16293-2422  & -             & \citet{Parise2004}\\
                                      &       & HH212            & $450$         & \citet{Bianchi2017b}\\
                                      &       & HH212            & $400$         & \citet{Taquet2019}\\
	                                    &       & Cep E-mm         & $730$         & \citet{Ospina-Zamudio2018}\\
                                      &       & NGC 7538-IRS1    & $2~800$       & \citet{Ospina-Zamudio2019a}\\
	                                    &       & NGC 1333-IRAS 4A & $293\pm22$    & \citet{Sahu2019}\\
	                                    &       & B1-c             & $320$         & \citet{vanGelder2020b}\\
	                                    &       & S68N             & $436$         & \citet{vanGelder2020b}\\
                                      &       & B1-bS            & $320$         & \citet{vanGelder2020b}\\
	                                    &       & G211.47-19.27S   & $415$         & \citet{Hsu2020}\\
                                      & $77$  & L1157-B1         & $250$         & \citet{Codella2012}\\
	                                    &       & AFGL5142-EC      & $1~800$       & \citet{Fontani2015}\\
	                                    &       & 05358-mm3        & $1~800$       & \citet{Fontani2015}\\
                                      &       & G034-G2(MM2)     & $2~900$       & \citet{Fontani2015}\\
                                      &       & G028-C1(MM9)     & $5~000$       & \citet{Fontani2015}\\
                                      &       & AFGL5142-MM      & $1~800$       & \citet{Fontani2015}\\
                                      &       & 05358-mm1        & $1~800$       & \citet{Fontani2015}\\
                                      &       & 18089-1732       & $3~600$       & \citet{Fontani2015}\\
                                      &       & 18517+0437       & $2~900$       & \citet{Fontani2015}\\
                                      &       & G75-core         & $3~800$       & \citet{Fontani2015}\\
                                      &       & I20293-MM1       & $2~000$       & \citet{Fontani2015}\\
	                                    &       & I23385           & $4~900$       & \citet{Fontani2015}\\
                                      &       & G5.89-0.39       & $1~280$       & \citet{Fontani2015}\\
                                      &       & I19035-VLA1      & $2~200$       & \citet{Fontani2015}\\
                                      &       & 19410+2336       & $2~100$       & \citet{Fontani2015}\\
                                      &       & ON1              & $2~500$       & \citet{Fontani2015}\\
                                      &       & 23033+5951       & $3~500$       & \citet{Fontani2015}\\
                                      &       & NGC7538-IRS9     & $2~800$       & \citet{Fontani2015}\B\\
 \hline
                 $^{16}$O/$^{18}$O~~  & $450$ & NGC 6334I        & $1~300$       & \citet{Bogelund2018}\T\\
                                      & $557$ & IRAS 16293-2422  & $140$         & \citet{Manigand2020a}\\
                                      & $560$ & IRAS 16293-2422  & -             & \citet{Jorgensen2018}\\
	                                    &       & B1-c             & $320$         & \citet{vanGelder2020b}\\
	                                    &       & S68N             & $436$         & \citet{vanGelder2020b}\\
	                                    &       & B1-bS            & $320$         & \citet{vanGelder2020b}\\
                                      &       & SMM1-a           & $436.0\pm9.2$ & Ligterink et al. 2020d in prep.\B\\
 \hline
 \end{xtabular*}
 \end{center}
\twocolumn}

\begin{landscape}
{\onecolumn
 \begin{adjustwidth}{-5cm}{}
 \topcaption{Column densities in cm$^{-2}$ of CH$_{3}$OH, CH$_{2}$DOH, and CH$_{3}$OD in star-forming regions and comets used to construct Figs.~\ref{fig_CH2DOH}--\ref{fig_Dmethtot} and~\ref{fig_CH2DOH_all}--\ref{fig_Dmethtot_all}. If a value appeared only in Figs.~\ref{fig_CH2DOH_all}--\ref{fig_Dmethtot_all} and not in Figs.~\ref{fig_CH2DOH}--\ref{fig_Dmethtot}, which contain only the most-reliable interstellar methanol observations, then the last column states the reason for that (as discussed in Section~\ref{Dmeth}). The full descriptions of the probed locations in each source, beam sizes, and source size assumptions can be found in the corresponding publications.}
 \label{tabl_Dmeth}
 \tablefirsthead{\hline \multicolumn{1}{l}{\textbf{Source}} & \multicolumn{1}{l}{\textbf{Position}} & \multicolumn{1}{l}{\textbf{CH$_{3}$OH}} & \multicolumn{1}{l}{\textbf{CH$_{2}$DOH}} & \multicolumn{1}{l}{\textbf{CH$_{3}$OD}} & \multicolumn{1}{l}{\textbf{Reference}} & \multicolumn{1}{l}{\textbf{Reason for Exclusion}} \\ \hline}
 \tablehead{\multicolumn{7}{c}%
           {{\captionsize\bfseries  \thetable{} -- continued from previous page}} \\
   \hline \multicolumn{1}{l}{\textbf{Source}} & \multicolumn{1}{l}{\textbf{Position}} & \multicolumn{1}{l}{\textbf{CH$_{3}$OH}} & \multicolumn{1}{l}{\textbf{CH$_{2}$DOH}} & \multicolumn{1}{l}{\textbf{CH$_{3}$OD}} & \multicolumn{1}{l}{\textbf{Reference}} & \multicolumn{1}{l}{\textbf{Reason for Exclusion}} \\ \hline}
 \tabletail{}
 \tablelasttail{}
 \begin{xtabular*}{1.3\textheight}{lllllll}
 L1544 & dust peak & $3.90 \times 10^{13}$ & $2.80 \times 10^{12}$ & - & \citet{Chacon-Tanarro2019} & -\T\\
   & methanol peak & $5.90 \times 10^{13}$ & $3.30 \times 10^{12}$ & - & \citet{Chacon-Tanarro2019} & -\\
   & dust peak & $2.70 \times 10^{13}$ & $2.40 \times 10^{12}$ & $<2.40 \times 10^{11}$ & \citet{Bizzocchi2014} & -\\
 L183 & dust peak & $4.90 \times 10^{13}$ & $1.90 \times 10^{12}$ & - & \citet{Lattanzi2020} & -\\
 L1495-B10 & Seo06 & $2.60 \times 10^{13}$ & $1.00 \times 10^{12}$ & - & Ambrose et al. 2020 subm. & -\\
  & Seo07 & $1.00 \times 10^{13}$ & $1.35 \times 10^{12}$ & - & Ambrose et al. 2020 subm. & -\\
  & Seo08 & $1.40 \times 10^{13}$ & $1.98 \times 10^{12}$ & - & Ambrose et al. 2020 subm. & -\\
  & Seo09 & $2.30 \times 10^{13}$ & $2.87 \times 10^{12}$ & - & Ambrose et al. 2020 subm. & -\\
  & Seo10 & $2.20 \times 10^{13}$ & $2.49 \times 10^{12}$ & - & Ambrose et al. 2020 subm. & -\\
  & Seo11 & $2.60 \times 10^{13}$ & $<1.37 \times 10^{12}$ & - & Ambrose et al. 2020 subm. & -\\
  & Seo12 & $2.90 \times 10^{13}$ & $1.63 \times 10^{12}$ & - & Ambrose et al. 2020 subm. & -\\
  & Seo13 & $2.50 \times 10^{13}$ & $<2.54 \times 10^{12}$ & - & Ambrose et al. 2020 subm. & -\\
  & Seo14 & $3.40 \times 10^{13}$ & $<1.26 \times 10^{12}$ & - & Ambrose et al. 2020 subm. & -\\
  & Seo15 & $1.50 \times 10^{13}$ & $1.57 \times 10^{12}$ & - & Ambrose et al. 2020 subm. & -\\
  & Seo16 & $1.60 \times 10^{13}$ & $1.26 \times 10^{12}$ & - & Ambrose et al. 2020 subm. & -\\
  & Seo17 & $8.40 \times 10^{12}$ & $1.90 \times 10^{12}$ & - & Ambrose et al. 2020 subm. & -\\
 I00117-MM2 & on source & $1.80 \times 10^{14}$ & $<1.30 \times 10^{12}$ & - & \citet{Fontani2015} & single dish\\
 AFGL5142-EC & on source & $6.15 \times 10^{15}$ & $1.10 \times 10^{13}$ & - & \citet{Fontani2015} & single dish\\
  & on source ($^{13}$CH$_{3}$OH) & $4.06 \times 10^{15}$ & $1.10 \times 10^{13}$ & - & \citet{Fontani2015} & single dish\\
 05358-mm3 & on source & $2.49 \times 10^{15}$ & $8.00 \times 10^{12}$ & - & \citet{Fontani2015} & single dish\\
  & on source ($^{13}$CH$_{3}$OH) & $1.18 \times 10^{15}$ & $8.00 \times 10^{12}$ & - & \citet{Fontani2015} & single dish\\
 G034-G2(MM2) & on source & $1.75 \times 10^{14}$ & $3.00 \times 10^{12}$ & - & \citet{Fontani2015} & single dish\\
  & on source ($^{13}$CH$_{3}$OH) & $6.93 \times 10^{13}$ & $3.00 \times 10^{12}$ & - & \citet{Fontani2015} & single dish\\
 G034-F2(MM7) & on source & $9.50 \times 10^{13}$ & $<7.00 \times 10^{11}$ & - & \citet{Fontani2015} & single dish\\
 G034-F1(MM8) & on source & $2.24 \times 10^{14}$ & $<1.00 \times 10^{12}$ & - & \citet{Fontani2015} & single dish\\
 G028-C1(MM9) & on source & $2.69 \times 10^{14}$ & $<1.00 \times 10^{12}$ & - & \citet{Fontani2015} & single dish\\
  & on source ($^{13}$CH$_{3}$OH) & $5.47 \times 10^{14}$ & $<1.00 \times 10^{12}$ & - & \citet{Fontani2015} & single dish\\
 I20293-WC & on source & $3.44 \times 10^{14}$ & $<2.00 \times 10^{12}$ & - & \citet{Fontani2015} & single dish\\
 I22134-G & on source & $2.87 \times 10^{14}$ & $<1.00 \times 10^{12}$ & - & \citet{Fontani2015} & single dish\\
 I22134-B & on source & $3.50 \times 10^{13}$ & $<7.00 \times 10^{11}$ & - & \citet{Fontani2015} & single dish\\
 L1157-B1 & shocked outer envelope ($^{13}$CH$_{3}$OH) & $2.31 \times 10^{15}$ & $4.00 \times 10^{13}$ & - & \citet{Codella2012} & -\\
 NGC 1333-IRAS 2A~~ & ellip. mask molec. peak (P04) & $5.00 \times 10^{18}$ & $2.90 \times 10^{17}$ & $7.90 \times 10^{16}$ & \citet{Taquet2019} & -\\
  & ellip. mask molec. peak (J18) & $5.00 \times 10^{18}$ & $2.90 \times 10^{17}$ & $3.60 \times 10^{17}$ & \citet{Taquet2019} & -\\
  & $10\arcsec$ on source & $1.01 \times 10^{15}$ & $5.20 \times 10^{14}$ & $<8.00 \times 10^{13}$ & \citet{Parise2006} & single dish\\
 NGC 1333-IRAS 4A~~ & ellip. mask molec. peak (P04) & $1.60 \times 10^{19}$ & $5.90 \times 10^{17}$ & $1.10 \times 10^{17}$ & \citet{Taquet2019} & -\\
  & ellip. mask molec. peak (J18) & $1.60 \times 10^{19}$ & $5.90 \times 10^{17}$ & $5.00 \times 10^{17}$ & \citet{Taquet2019} & -\\
  & $10\arcsec$ on source & $6.90 \times 10^{14}$ & $4.30 \times 10^{14}$ & $3.10 \times 10^{13}$ & \citet{Parise2006} & single dish\\
  & A2 ($^{13}$CH$_{3}$OH) & $2.25 \times 10^{19}$ & $1.30 \times 10^{17}$ & - & \citet{Sahu2019} & -\\
  & A1 ($^{13}$CH$_{3}$OH) & $1.34 \times 10^{17}$ & $6.51 \times 10^{15}$ & - & \citet{Sahu2019} & -\\
 HH212 & ellip. mask molec. peak (P04; $^{13}$CH$_{3}$OH) & $2.24 \times 10^{18}$ & $6.40 \times 10^{16}$ & $9.90 \times 10^{15}$ & \citet{Taquet2019} & -\\
  & ellip. mask molec. peak (J18; $^{13}$CH$_{3}$OH) & $2.24 \times 10^{18}$ & $6.40 \times 10^{16}$ & $4.40 \times 10^{16}$ & \citet{Taquet2019} & -\\
  & continuum peak inner $100$~au ($^{13}$CH$_{3}$OH) & $4.55 \times 10^{18}$ & $1.10 \times 10^{17}$ & - & \citet{Bianchi2017b} & -\\
  & lower disc atmosphere ($^{13}$CH$_{3}$OH) & $1.25 \times 10^{18}$ & $1.60 \times 10^{17}$ & - & \citet{LeeChin-Fei2019a} & -\\
  & lower disc atmosphere & $3.40 \times 10^{17}$ & $9.20 \times 10^{16}$ & - & \citet{LeeChin-Fei2017b} & -\\
 IRAS 16293-2422 & $10\arcsec$ circumbinary env. $20$~K & $3.50 \times 10^{15}$ & $3.00 \times 10^{15}$ & $1.50 \times 10^{14}$ & \citet{Parise2002} & single dish\\
  & $10\arcsec$ circumbinary env. $48$~K & $3.50 \times 10^{15}$ & $3.00 \times 10^{15}$ & $2.80 \times 10^{14}$ & \citet{Parise2002} & single dish\\
  & $10\arcsec$ circumbinary env. ($^{13}$CH$_{3}$OH) & $9.80 \times 10^{15}$ & $3.00 \times 10^{15}$ & $1.50 \times 10^{14}$ & \citet{Parise2004} & single dish\\
  & $0.6\arcsec$ offset NE from A & $1.30 \times 10^{19}$ & $1.10 \times 10^{18}$ & $2.80 \times 10^{17}$ & \citet{Manigand2020a} & -\\
  & $0.6\arcsec$ offset NE from A ($^{13}$CH$_{3}$OH) & $1.36 \times 10^{19}$ & $1.10 \times 10^{18}$ & $2.80 \times 10^{17}$ & \citet{Manigand2020a} & -\\
  & $0.6\arcsec$ offset NE from A (CH$_{3}^{18}$OH) & $1.28 \times 10^{19}$ & $1.10 \times 10^{18}$ & $2.80 \times 10^{17}$ & \citet{Manigand2020a} & -\\
  & $0.5\arcsec$ offset SW from B (CH$_{3}^{18}$OH) & $1.00 \times 10^{19}$ & $7.10 \times 10^{17}$ & $1.80 \times 10^{17}$ & \citet{Jorgensen2018} & -\T\\
  & $20\arcsec$ circumbinary env. & $8.80 \times 10^{14}$ & - & $<8.00 \times 10^{12}$ & \citet{vanDishoeck1995} & single dish\\
 NGC 1333-IRAS 4B~~ & $10\arcsec$ on source & $8.00 \times 10^{14}$ & $2.90 \times 10^{14}$ & $1.10 \times 10^{13}$ & \citet{Parise2006} & single dish\\
 L1448N & $10\arcsec$ on source & $1.20 \times 10^{14}$ & $2.10 \times 10^{14}$ & $<8.00 \times 10^{13}$ & \citet{Parise2006} & single dish\\
 L1448mm & $10\arcsec$ on source & $1.60 \times 10^{14}$ & $<1.13 \times 10^{15}$ & $<4.00 \times 10^{14}$ & \citet{Parise2006} & single dish\\
 L1157mm & $10\arcsec$ on source & $1.90 \times 10^{14}$ & $<1.03 \times 10^{15}$ & $<2.20 \times 10^{14}$ & \citet{Parise2006} & single dish\\
 SVS13-A & low-T $3\arcsec$ component ($^{13}$CH$_{3}$OH) & $1.09 \times 10^{17}$ & $7.00 \times 10^{14}$ & $6.00 \times 10^{14}$ & \citet{Bianchi2017a} & -\\
  & high-T $0.3\arcsec$ component ($^{13}$CH$_{3}$OH) & $1.36 \times 10^{19}$ & $4.00 \times 10^{17}$ & - & \citet{Bianchi2017a} & -\\
 L483 & on source ($^{13}$CH$_{3}$OH) & $2.92 \times 10^{14}$ & $5.50 \times 10^{12}$ & $4.00 \times 10^{12}$ & \citet{Agundez2019} & -\\
 L1527 & on source & $6.30 \times 10^{13}$ & $<1.90 \times 10^{12}$ & - & \citet{Sakai2009b} & -\\
 G211.47-19.27S & on source & $8.50 \times 10^{16}$ & $2.30 \times 10^{16}$ & - & \citet{Hsu2020} & optically thick CH$_{3}$OH\\
  & on source ($^{13}$CH$_{3}$OH) & $6.44 \times 10^{17}$ & $2.30 \times 10^{16}$ & - & \citet{Hsu2020} & -\\
 B1-c & Band 3 ($^{13}$CH$_{3}$OH) & $1.80 \times 10^{18}$ & $1.03 \times 10^{17}$ & - & \citet{vanGelder2020b} & -\\
  & Band 3 (CH$_{3}^{18}$OH) & $<1.80 \times 10^{16}$ & $1.03 \times 10^{17}$ & - & \citet{vanGelder2020b} & upper limit CH$_{3}^{18}$OH\\
  & Band 6 ($^{13}$CH$_{3}$OH) & $1.90 \times 10^{16}$ & $1.60 \times 10^{17}$ & - & \citet{vanGelder2020b} & optically thick $^{13}$CH$_{3}$OH\\
  & Band 6 (CH$_{3}^{18}$OH) & $1.90 \times 10^{18}$ & $1.60 \times 10^{17}$ & - & \citet{vanGelder2020b} & -\\
  & warm component ($^{13}$CH$_{3}$OH) & $1.26 \times 10^{18}$ & $1.60 \times 10^{17}$ & - & \citet{vanGelder2020b} & -\\
  & warm component (CH$_{3}^{18}$OH) & $1.90 \times 10^{18}$ & $1.60 \times 10^{17}$ & - & \citet{vanGelder2020b} & -\\
  & cold component ($^{13}$CH$_{3}$OH) & $1.19 \times 10^{16}$ & $<1.60 \times 10^{14}$ & - & \citet{vanGelder2020b} & -\\
 S68N & Band 3 ($^{13}$CH$_{3}$OH) & $3.60 \times 10^{17}$ & $<4.68 \times 10^{16}$ & - & \citet{vanGelder2020b} & -\\
  & Band 6 ($^{13}$CH$_{3}$OH) & $9.80 \times 10^{15}$ & $6.02 \times 10^{16}$ & - & \citet{vanGelder2020b} & optically thick $^{13}$CH$_{3}$OH\\
  & Band 6 (CH$_{3}^{18}$OH) & $1.40 \times 10^{18}$ & $6.02 \times 10^{16}$ & - & \citet{vanGelder2020b} & -\\
  & warm component ($^{13}$CH$_{3}$OH) & $7.00 \times 10^{17}$ & $6.00 \times 10^{16}$ & - & \citet{vanGelder2020b} & -\\
  & warm component (CH$_{3}^{18}$OH)  & $1.40 \times 10^{18}$ & $6.00 \times 10^{16}$ & - & \citet{vanGelder2020b} & -\\
 B1-bS & Band 3 ($^{13}$CH$_{3}$OH) & $2.40 \times 10^{17}$ & $<2.38 \times 10^{16}$ & - & \citet{vanGelder2020b} & -\\
  & Band 6 ($^{13}$CH$_{3}$OH) & $1.50 \times 10^{15}$ & $<2.35 \times 10^{16}$ & - & \citet{vanGelder2020b} & -\\
  & Band 6 (CH$_{3}^{18}$OH) & $5.00 \times 10^{17}$ & $<2.35 \times 10^{16}$ & - & \citet{vanGelder2020b} & -\\
 NGC 7129 FIRS 2 & compact hot core ($^{13}$CH$_{3}$OH) & $3.44 \times 10^{20}$ & $1.40 \times 10^{16}$ & - & \citet{Fuente2014} & -\\
 Cep E-mm & CepE-A ($^{13}$CH$_{3}$OH) & $4.90 \times 10^{17}$ & $1.77 \times 10^{16}$ & - & \citet{Ospina-Zamudio2018} & -\\
 SMM1-a & on source (CH$_{3}^{18}$OH) & $7.84 \times 10^{17}$ & $2.30 \times 10^{16}$ & - & Ligterink et al. 2020d in prep. & -\\
  &  on source ($^{13}$CH$_{3}$OH) & $5.30 \times 10^{17}$ & $2.30 \times 10^{16}$ & - & Ligterink et al. 2020d in prep. & -\\
 Orion KL & compact ridge ($^{13}$CH$_{3}$OH) & $6.00 \times 10^{17}$ & $3.50 \times 10^{15}$ & $3.00 \times 10^{15}$ & \citet{Neill2013b} & -\\
  & hot core ($^{13}$CH$_{3}$OH) & $6.60 \times 10^{17}$ & $<2.80 \times 10^{15}$ & $<1.20 \times 10^{15}$ & \citet{Neill2013b} & -\\
  & dM-1 $T_{\text{rot}}(\text{HCOOCH}_{3})$ & $4.20 \times 10^{18}$ & $4.70 \times 10^{15}$ & - & \citet{Peng2012} & -\\
  & dM-1 & $2.10 \times 10^{18}$ & $2.40 \times 10^{15}$ & - & \citet{Peng2012} & -\\
  & dM-2 $T_{\text{rot}}(\text{HCOOCH}_{3})$ & $2.20 \times 10^{18}$ & $1.70 \times 10^{15}$ & - & \citet{Peng2012} & -\\
  & dM-2 & $1.90 \times 10^{18}$ & $1.40 \times 10^{15}$ & - & \citet{Peng2012} & -\\
  & dM-3 & $1.70 \times 10^{18}$ & $1.50 \times 10^{15}$ & - & \citet{Peng2012} & -\\
  & KL-W & $9.20 \times 10^{17}$ & $<2.00 \times 10^{14}$ & - & \citet{Peng2012} & -\\
  & IRc2 & $4.70 \times 10^{17}$ & $8.00 \times 10^{14}$ & $4.40 \times 10^{15}$ & \citet{Peng2012} & -\\
 NGC 6334I & MM1 I ($^{13}$CH$_{3}$OH) & $5.15 \times 10^{19}$ & $1.10 \times 10^{17}$ & $5.50 \times 10^{17}$ & \citet{Bogelund2018} & -\\
  & MM1 I (CH$_{3}^{18}$OH) & $1.22 \times 10^{20}$ & $1.10 \times 10^{17}$ & $5.50 \times 10^{17}$ & \citet{Bogelund2018} & -\\
  & MM1 II ($^{13}$CH$_{3}$OH) & $4.59 \times 10^{19}$ & $5.20 \times 10^{16}$ & $3.80 \times 10^{17}$ & \citet{Bogelund2018} & -\\
  & MM1 II (CH$_{3}^{18}$OH) & $1.08 \times 10^{20}$ & $5.20 \times 10^{16}$ & $3.80 \times 10^{17}$ & \citet{Bogelund2018} & -\\
  & MM1 III ($^{13}$CH$_{3}$OH) & $5.15 \times 10^{19}$ & $7.40 \times 10^{16}$ & $2.30 \times 10^{17}$ & \citet{Bogelund2018} & -\\
  & MM1 III (CH$_{3}^{18}$OH) & $7.65 \times 10^{19}$ & $7.40 \times 10^{16}$ & $2.30 \times 10^{17}$ & \citet{Bogelund2018} & -\\
  & MM1 IV ($^{13}$CH$_{3}$OH) & $3.22 \times 10^{19}$ & $3.15 \times 10^{16}$ & $1.70 \times 10^{17}$ & \citet{Bogelund2018} & -\\
  & MM1 IV (CH$_{3}^{18}$OH) & $6.75 \times 10^{19}$ & $3.15 \times 10^{16}$ & $1.70 \times 10^{17}$ & \citet{Bogelund2018} & -\\
  & MM1 V ($^{13}$CH$_{3}$OH) & $8.06 \times 10^{18}$ & $1.25 \times 10^{16}$ & $4.30 \times 10^{16}$ & \citet{Bogelund2018} & -\\
  & MM1 V (CH$_{3}^{18}$OH) & $1.84 \times 10^{19}$ & $1.25 \times 10^{16}$ & $4.30 \times 10^{16}$ & \citet{Bogelund2018} & -\\
  & MM2 I ($^{13}$CH$_{3}$OH) & $4.09 \times 10^{19}$ & $9.15 \times 10^{16}$ & $1.80 \times 10^{17}$ & \citet{Bogelund2018} & -\\
  & MM2 I (CH$_{3}^{18}$OH) & $5.18 \times 10^{19}$ & $9.15 \times 10^{16}$ & $1.80 \times 10^{17}$ & \citet{Bogelund2018} & -\\
  & MM2 II ($^{13}$CH$_{3}$OH) & $1.12 \times 10^{19}$ & $6.50 \times 10^{15}$ & $4.00 \times 10^{16}$ & \citet{Bogelund2018} & -\\
  & MM2 II (CH$_{3}^{18}$OH) & $1.44 \times 10^{19}$ & $6.50 \times 10^{15}$ & $4.00 \times 10^{16}$ & \citet{Bogelund2018} & -\\
  & MM3 I ($^{13}$CH$_{3}$OH) & $5.58 \times 10^{18}$ & $4.50 \times 10^{15}$ & $1.50 \times 10^{16}$ & \citet{Bogelund2018} & -\\
  & MM3 I (CH$_{3}^{18}$OH) & $6.30 \times 10^{18}$ & $4.50 \times 10^{15}$ & $1.50 \times 10^{16}$ & \citet{Bogelund2018} & -\\
  & MM3 II ($^{13}$CH$_{3}$OH) & $4.96 \times 10^{18}$ & $4.50 \times 10^{15}$ & $1.60 \times 10^{16}$ & \citet{Bogelund2018} & -\\
  & MM3 II (CH$_{3}^{18}$OH) & $5.85 \times 10^{18}$ & $4.50 \times 10^{15}$ & $1.60 \times 10^{16}$ & \citet{Bogelund2018} & -\\
 RAFGL 7009S & $8\pm5\arcsec$ on source & $1.20 \times 10^{16}$ & - & $9.50 \times 10^{13}$ & \citet{Dartois2000} & -\\
 Sgr B2(N2) & $2\arcsec$ on source & $4.00 \times 10^{19}$ & $4.80 \times 10^{16}$ & $<2.60 \times 10^{16}$ & \citet{Belloche2016} & -\\
 NGC 7538-IRS1 & CepE-A ($^{13}$CH$_{3}$OH) & $3.78 \times 10^{17}$ & $1.20 \times 10^{15}$ & $3.80 \times 10^{14}$ & \citet{Ospina-Zamudio2019a} & single dish\\
 I00117-MM1 & on source & $1.22 \times 10^{14}$ & $<2.00 \times 10^{12}$ & - & \citet{Fontani2015} & single dish\\
 AFGL5142-MM & on source & $2.63 \times 10^{16}$ & $1.06 \times 10^{15}$ & - & \citet{Fontani2015} & single dish\\
  & on source ($^{13}$CH$_{3}$OH) & $4.47 \times 10^{15}$ & $1.06 \times 10^{15}$ & - & \citet{Fontani2015} & single dish\\
 05358-mm1 & on source & $1.25 \times 10^{16}$ & $<1.30 \times 10^{13}$ & - & \citet{Fontani2015} & single dish\\
  & on source ($^{13}$CH$_{3}$OH) & $4.70 \times 10^{15}$ & $<1.30 \times 10^{13}$ & - & \citet{Fontani2015} & single dish\\
 18089-1732 & on source & $3.18 \times 10^{16}$ & $9.00 \times 10^{14}$ & - & \citet{Fontani2015} & single dish\\
  & on source ($^{13}$CH$_{3}$OH) & $4.94 \times 10^{16}$ & $9.00 \times 10^{14}$ & - & \citet{Fontani2015} & single dish\T\\
 18517+0437 & on source & $2.09 \times 10^{16}$ & $<2.00 \times 10^{13}$ & - & \citet{Fontani2015} & single dish\\
  & on source ($^{13}$CH$_{3}$OH) & $1.97 \times 10^{16}$ & $<2.00 \times 10^{13}$ & - & \citet{Fontani2015} & single dish\\
 G75-core & on source & $1.51 \times 10^{16}$ & $5.50 \times 10^{13}$ & - & \citet{Fontani2015} & single dish\\
  & on source ($^{13}$CH$_{3}$OH) & $3.23 \times 10^{15}$ & $5.50 \times 10^{13}$ & - & \citet{Fontani2015} & single dish\\
 I20293-MM1 & on source & $2.75 \times 10^{15}$ & $<4.00 \times 10^{12}$ & - & \citet{Fontani2015} & single dish\\
  & on source ($^{13}$CH$_{3}$OH) & $6.93 \times 10^{14}$ & $<4.00 \times 10^{12}$ & - & \citet{Fontani2015} & single dish\\
 I21307 & on source & $6.54 \times 10^{14}$ & $<3.00 \times 10^{12}$ & - & \citet{Fontani2015} & single dish\\
 I23385 & on source & $1.80 \times 10^{15}$ & $<2.00 \times 10^{12}$ & - & \citet{Fontani2015} & single dish\\
  & on source ($^{13}$CH$_{3}$OH) & $2.31 \times 10^{14}$ & $<2.00 \times 10^{12}$ & - & \citet{Fontani2015} & single dish\\
 G5.89-0.39 & on source & $1.28 \times 10^{16}$ & $<1.40 \times 10^{13}$ & - & \citet{Fontani2015} & single dish\\
  & on source ($^{13}$CH$_{3}$OH) & $1.08 \times 10^{16}$ & $<1.40 \times 10^{13}$ & - & \citet{Fontani2015} & single dish\\
 I19035-VLA1 & on source & $1.64 \times 10^{15}$ & $<3.00 \times 10^{12}$ & - & \citet{Fontani2015} & single dish\\
  & on source ($^{13}$CH$_{3}$OH) & $3.70 \times 10^{15}$ & $<3.00 \times 10^{12}$ & - & \citet{Fontani2015} & single dish\\
 19410+2336 & on source & $2.02 \times 10^{15}$ & $<3.00 \times 10^{12}$ & - & \citet{Fontani2015} & single dish\\
  & on source ($^{13}$CH$_{3}$OH) & $4.46 \times 10^{15}$ & $<3.00 \times 10^{12}$ & - & \citet{Fontani2015} & single dish\\
 ON1 & on source & $3.24 \times 10^{15}$ & $<2.00 \times 10^{12}$ & - & \citet{Fontani2015} & single dish\\
  & on source ($^{13}$CH$_{3}$OH) & $6.91 \times 10^{15}$ & $<2.00 \times 10^{12}$ & - & \citet{Fontani2015} & single dish\\
 I22134-VLA1 & on source & $1.64 \times 10^{14}$ & $<2.00 \times 10^{12}$ & - & \citet{Fontani2015} & single dish\\
 23033+5951 & on source & $1.20 \times 10^{15}$ & $<2.00 \times 10^{12}$ & - & \citet{Fontani2015} & single dish\\
  & on source ($^{13}$CH$_{3}$OH) & $6.41 \times 10^{15}$ & $<2.00 \times 10^{12}$ & - & \citet{Fontani2015} & single dish\\
 NGC7538-IRS9 & on source & $1.76 \times 10^{15}$ & $<2.00 \times 10^{12}$ & - & \citet{Fontani2015} & single dish\\
  & on source ($^{13}$CH$_{3}$OH) & $3.85 \times 10^{14}$ & $<2.00 \times 10^{12}$ & - & \citet{Fontani2015} & single dish\\
 C/1995 O1 (Hale--Bopp)~~ & cometary coma & $2.40 \times 10^{-2}$ & $<6.00 \times 10^{-4}$ & $<7.00 \times 10^{-4}$ & \citet{Crovisier2004} & -\\
 \end{xtabular*}
 \end{adjustwidth}
\twocolumn
}
\end{landscape}

\bsp 

\label{lastpage}

\end{document}